\documentclass[aps, pre, amsfonts, amssymb, amsmath, showpacs, reprint]{revtex4-1}
\usepackage{times}
\usepackage{bm}
\usepackage{hyperref}
\usepackage{graphicx}
\usepackage{mathtools}
\usepackage{epsfig} 
\usepackage{color}
\usepackage{float}
\usepackage{hyperref}
\usepackage{cleveref}

\begin{document}

\title{Impact of micro-scale stochastic Zonal Flows on the macro-scale V-RMHD modes}
\author{Sara Moradi,}
\email{sara.moradi@ukaea.uk}
\affiliation{Laboratory for Plasma Physics - LPP-ERM/KMS, Royal Military Academy, 1000-Brussels, Belgium}
\author{A. Thyagaraja,} 
\affiliation{Astrophysics Group, Dept. of Physics, University of Bristol, Bristol, BS8,1TL, United Kingdom}
%
%
%

\begin{abstract}
A model is developed to simulate micro-scale turbulence driven ZFs, and their impact on the MHD tearing and kink modes is examined. The model is based on a stochastic representation of the micro-scale ZFs with a given Alfv\'{e}n Mach number, $M_S$. Two approaches were explored: i) passive stochastic model where the ZFs amplitudes are independent of the MHD mode amplitude, and ii) the semi-stochastic model where the amplitudes of the ZFs have a dependence on the amplitude of the MHD mode itself. The results show that the stochastic ZFs can significantly stabilise the (2,1) and (1,1) MHD modes even at very low kinematic viscosity, $Pr$, where the mode is linearly unstable. Our results, therefore indicate a possible mechanism for stabilisation of the MHD modes via small-scale perturbations in poloidal flow, simulating the turbulence driven ZFs. 
\end{abstract}
\pacs{}
\maketitle
\section{Introduction} 
The development of large scale (i.e. machine size) Magneto-Hydro-Dynamic (MHD) instabilities such as $(m=1/n=1)$ internal kink mode \cite{Marshall1973,VonGoeler1974,Ksdomtsev1975,Thyagaraja1991a,Porcelli1996,Monticello1986,Mendonca2018,Kai-Qi2018} and/or $(m=2/n=1)$, Tearing Modes (TM) of which a special case for fusion plasmas is the Neoclassical-Tearing Mode \cite{Wilson1996,Haye2002,Smolyakov1993,Hegna1998,ZChang1995,Carrera1986,McDevitt2006,Chandra2015} are important limiting factors for establishing a stable confined plasma for long enough time to reach the fusion conditions. 

The $(1,1)$ kink mode arises inside the $q=1$ rational surface when $q$ at the axis is $<1$, where $q$ is the safety factor. This mode is observed to trigger the sawtooth oscillations in the plasma core, influencing the confinement quality. Most notable impact of the sawtooth oscillation, is a slow rise of the order of the resistive time scale, in the plasma temperature in the centre, followed by an abrupt crash (on a 100 microsecond time scale), relaxing the temperature profiles back to the initial values at the beginning of each cycle \cite{VonGoeler1974,Larche2017}. The loss of temperature in the central plasma region limits the achievable pressure profile. However, the sawtooth crash benefits the plasma core confinement by removing the unwanted impurities from the core. Control of the sawtooth crashes e.g. their amplitude and frequency, hence would be important for optimising plasma performance in a fusion device \cite{Campbell1988,McClements1996,Larche2017,Yang2012,Mao2001}.

The classical $(2,1)$ TM mode can result in the destruction of the topology of the nested magnetic flux surfaces, where the reconnection of the field lines produces magnetic islands in the confined region. The continuous growth of the islands then destroys magnetic confinement and can produce runaway electrons, and lead to major disruptions. A disruption is an event that induces a rapid loss of thermal energy followed by a quench of the plasma current. Disruptions release large heat loads and electro-magnetic forces on the surrounding machine structures, and can result in a significant damage to the Plasma Facing Components (PFCs). 

When the classical $(2,1)$ TM is linearly {\it stable} (when $\Delta'< 0$), it was shown both theoretically and experimentally \cite{Wilson1996,Hegna1998,DeVries1997}, that Neoclassical bootstrap currents (or their absence in the interior of a magnetic island) can lead to a non-linear destabilisation of the island resulting in NTMs. This is a non-linear instability which requires an initial seed island of sufficient threshold amplitude. 

Given the important consequences of the MHD instabilities (both linear and non-linear) which generate global deleterious effects, understanding their dynamics and the development of control actuators to avoid them is crucial for current and future fusion machines \cite{DeVries2009,DeVries2014,Lehnen2013}.

Another limiting factor for achieving fusion is the confinement time. This is limited by the existence of a zoo of small-scale electro-magnetic turbulence (due to the ion/electron drift waves) which are ubiquitous in magnetically confined fusion plasmas, and are considered to be responsible for anomalous transport. Although the small-scale turbulence does not result directly in a disruption, it can modify/enhance the large scale MHD modes through non-linear mode coupling and/or the generation of sheared Zonal Flows (ZFs). In a recent review by Ishizawa and co-authors \cite{Ishizawa2019}, the findings of an extensive body of work that has been devoted to the study of the multi-scale interactions and the interplay between turbulence and large scale TMs are examined. We will not try to re-examine these works here and will only refer to the a summary: the inverse energy cascade via non-linear parity mixing of the small-scale ballooning modes (such as ITG, ETG, KBM with twisting parity) \cite{Ishizawa2015,Coppi1977,Kadomtsev1971,Dorland2000,Pueschel2008} or electron Micro-TMs (with tearing parity) \cite{Doerk2011,Guttenfolder2012,moradi2013}, is believed to result in the destabilisation of otherwise stable TMs (e.g. reduce the stability threshold as compared to purely MHD analysis) or further growth of the unstable TMs (i.e. enhancement of the growth rate of the island). Thus, turbulence can create large-scale magnetic islands by the merging of small-scale islands. Other mechanisms such as anomalous current drive by turbulence i.e. much like a small-scale dynamo effect \cite{Li2012}, or turbulence-driven polarisation current as the result of turbulence-generated ZFs following the reconnected field lines of the island \cite{Ishizawa2013}, have also been suggested as drivers of magnetic seed islands. 

The back reaction of the magnetic islands on the turbulence is expressed through a direct energy cascade from MHD modes to turbulence as a result of the non-linear mode coupling observed in EAST tokamak \cite{Sun2018}. Moreover the magnetic perturbation of the field lines, especially around the X-point of the island increases the turbulence outside the island \cite{DeVries2014}, while the coherent vortex flows generated by the interactions between the turbulence and MHD mode at the boundary can act as a transport barrier that leads to suppression of the drift-wave instabilities inside the island \cite{Li2009,Ishizawa2013,poli2009,harriri2015,hornsby2015,Bardoczi2017}. The enhancement of the turbulence outside, and its reduction inside the island, along with the parallel transport results in increased flux which flattens (not completely) the temperature profile inside the separatrix of the island. This flattening of the temperature profile, changes the resistivity inside the island and further reduces the perturbed bootstrap current which has a destabilising effect on the island. It increases the growth of the island and eventually the degradation of the confinement outside the island equilibrates the profiles inside and outside leading to its saturation at lower plasma stored energy \cite{Bardoczi2016,Bardoczi2017}. 

It is a common view that in fusion plasmas turbulence is regulated by its self generating sheared $E\times B$ flows (ZFs) \cite{Diamond2005,Biskamp1983,Diamond1984}. The mechanism is explained as generation of the sheared flows via non-linear interactions producing Reynolds-Stresses, which in turn result in shearing of the turbulent eddies, thus reducing their radial correlation length \cite{Ham1995, Ham2015}. However, it is argued that in high $\beta$ plasma regimes, where $\beta$ is the ratio of plasma kinetic energy to magnetic energy, the ITGs responsible for generating the ZFs are stabilised due to magnetic field bending at high $\beta$, and KBM, TEM or ETG modes arise which do not produce strong ZFs. These micro-scale modes can grow beyond a physically relevant saturation level, unless through mode coupling they transfer their energy to stable modes \cite{Ishizawa2015} or form coherent structures such as magnetic islands \cite{Ishizawa2019}. 

Given the important role of the small-scale turbulence on the stability regime of the large scale MHD modes and their regulation by ZFs, in a more general approach, here we examine the impact of a stochastic ZF following a given probability distribution function (PDF) on the large-scale (1,1) and (2,1) modes. Our aim in this paper is to consider the possibility of dynamic {\it stabilisation} of the MHD modes by the stochastic ZFs. We note that it is well-known theoretically and experimentally \cite{Thyagaraja1991,Thyagaraja1993,Yang2012,Mao2001}, that in principle, one can dynamically stabilise MHD modes using external applied electro-magnetic perturbations with specific frequencies and amplitudes. We further extend this idea by including a spectrum of random perturbations on small-scales and show that in principle, their associated ZFs can result in stabilisation of the linear and non-linear unstable (1,1) and (2,1) visco-resistive MHD modes. 

Our results are based on a series of numerical simulations with the visco-resistive MHD version of the CUTIE code \cite{Mendonca2018,Chandra2015} where a stochastic ZFs model has been included in the vorticity equation. CUTIE is a non-linear, global, electromagnetic, quasi-neutral, visco-resistive MHD code in a large aspect-ratio cylindrical geometry (neglecting linear toroidal coupling due to curvature effects while including non-linear mode interactions). 

Two models were considered: i) passive stochastic ZFs model where the amplitudes of the ZFs are independent of the amplitude of the MHD mode; ii) semi-stochastic coupling model where the amplitudes of the ZFs are proportional to the amplitude of the mode itself. The first model aims to examine the impact of a passive stochastic flow arising from the small-scale turbulence on the MHD stability. The second model aims to represent, in a general manner, the interplay between the MHD mode and the turbulence generated stochastic flows following a direct-cascade phenomenology where the energy is transferred from the large scales to the small-scales which back-react with the mode itself. Our results show that above a certain critical amplitude, the stochastic ZFs have a strong stabilising effect on the (1,1) and (2,1) modes, even at a very low kinematic viscosity (e.g. $Pr=1$) where the modes are {\it linearly} unstable.   

The paper is arranged as follows: In section II we describe the details of our simple model and the CUTIE code. In section III we present the results of the stochastic passive model or the (2,1) mode. In section IV the results for both (1,1) and (2,1) modes are shown after applying the semi-stochastic ZFs model. Section V contains a summary of our findings and our conclusions.
\section{Model description} 
The CUTIE code employes a visco-resistive MHD model of plasma in a periodic cylindrical geometry $(r, \theta, z)$, where $r$ is the radial, $\theta$ is the azimuthal, and $z$ is the axial coordinate \cite{Chandra2015,Mendonca2018}. The equations for the mean and fluctuating parts of the electro-magnetic fields are split using a Fourier expansion where the mean fields are represented by the $(0,0)$ components, and the remaining terms represent the fluctuating components of the fields. The mean fields are co-evolved with the equations for fluctuating quantities. The fluctuation equations for the vorticity $\tilde{W}$ and poloidal flux function $\tilde{\psi}$ are:

\begin{eqnarray}
\frac{\partial \tilde{W}}{\partial t} + \mathbf{v}_0 \cdot \nabla \tilde{W} + v_{A}\nabla_{\parallel} \rho_{s}^2 \nabla_{\perp}^{2} \bar{\psi}\nonumber\\
= v_{A}\rho_s \frac{1}{r} \frac{\partial \tilde{\psi}}{\partial \theta} \frac{4\pi\rho_s}{cB_0}j^{'}_{0} + \frac{v_{th}\rho_s}{r}\{\tilde{\psi}, \rho_s^2\nabla_{\perp}^2\tilde{\psi}\}\nonumber\\
+ \frac{v_{th}\rho_s}{r}\{\tilde{W},\tilde{\phi}\} + \frac{\rho_s^2 }{r} \frac{d W_{S}}{dr}\frac{\partial \tilde{\phi}}{\partial \theta} + \nu\nabla_{\perp}^2 \tilde{W}, \label{1}\nonumber\\ 
\\
\frac{\partial\tilde{\psi}}{\partial t} + \mathbf{v}_{0}\cdot\nabla\tilde{\psi} + v_{A} \nabla_{\parallel}\tilde{\phi} = \frac{v_{th}\rho_s}{r}\{\tilde{\psi},\tilde{\phi}\} + \frac{c^2\eta}{4\pi}\nabla_{\perp}^2\tilde{\psi}, \label{2}\nonumber\\
\end{eqnarray}
where vorticity is defined as: 
\begin{equation}
\tilde{W} = \rho_s^2  \nabla_{\perp}^2\tilde{\phi}  \label{3}
\end{equation}
and the vorticity due to the stochastic ZFs is defined as
\begin{equation}
W_S =  \nabla\times (\delta v_s \mathbf{e}_{\theta}) = \frac{1}{r}  \frac{d}{d r} (r \delta v_s) \label{4}
\end{equation}

In the present problem $T_{0i}=T_{0e}=0.275$keV, and $n_0=4\times 10^{14}$cm$^{-3}$ are considered uniform and constant, with $T_{0i}$ and $T_{0e}$ being the ion and electron temperatures, and $n_0$ is the mean density, respectively. Here, $\rho = r/a$ following the \cite{Mendonca2018}, is assumed as $a=10$cm and $R_0 = 100$cm; and in this large aspect ratio limit the linear curvature effects are neglected. The resistivity $\eta$ and viscosity $\nu$ are {\it specified quantities} and are kept constant during the simulations. Additionally, $\rho_s = \frac{v_{th}}{\omega_{ci}}$, where $v_{th}^2 = (T_{0i} + T_{0e})/m_i$, $\omega_{ci} = (eB_0/m_ic)$, $e$ is the elementary charge and $m_i$ is the ion mass.  $v_A = B_0/(4\pi m_i n_0)^{1/2} = 2.18\times 10^8$cm/s is the Alfv\'{e}n velocity, and $\mathbf{v}_0$ is the sub-Alfv\'{e}nic equilibrium sheared flow (both axial and poloidal contributions) which is set to zero in the current study. The equilibrium current density is defined as $j_0=-\frac{c}{4\pi}\frac{1}{r}\frac{d}{dr}(r\frac{d\psi_0}{dr})$, where $\psi_0$ is the poloidal equilibrium flux function. 

Alfv\'{e}n time is set as $\tau_A= a/v_A$, and resistive and viscous time are defined as: $\tau_{\eta} = \frac{4\pi a^2}{c^2\eta}$, and $\tau_{\nu} = \frac{a^2}{\nu}$. Prandtl number is $Pr =\frac{\tau_{\eta}}{\tau_{\nu}}$, and the Lundquist number is $S = \frac{\tau_{\eta}}{\tau_{A}}$, and the corresponding Reynolds number is defined as: $Re = \frac{\tau_{\nu}}{\tau_A}$. $B_0 = 2T$, and $S=10^{6}$ are assumed. The profile of the safety-factor, $q$ ($=rB_z/RB_{0\theta}$ where $B_z=B_0$, and $B_{0\theta}= -d\psi_0/dr$), is prescribed as function of $\rho$: $q = q_0 [1+(\rho/\rho_0)^{2\lambda}]^{\lambda}$ with $[q_0 = 0.9, \lambda =1, \rho_0 = 0.6] $, and $[q_0 = 1.4, \lambda=2, \rho_0 = 0.74]$ for the (1,1) and (2,1) cases, respectively \cite{Mendonca2018,Chandra2015}. Since we prescribe both $B_0$, $S$ and $q$, the profile of $\eta$ is obtained from the condition that $\eta j_0$ is constant.

In the simulations presented we have used 101 radial grid points, 17 poloidal and 9 toroidal Fourier modes. The time step used in the simulations was $\Delta t  = 2\times 10^{-10}s$, thus, $v_A\Delta t / \Delta r = 0.44$ is the radial Courant number.

In the case of a passive stochastic ZFs model, the ZF velocity $\delta v_S$ is defined as follows: $\delta v_S = M_S X(\rho,t) v_A$ where $M_S$ is specified to be the stochastic ZFs Alfv\'{e}n Mach number, and $X(\rho,t)$ is the stochastic variable is assumed as a white gaussian noise, with zero mean and variance of 1. Here, $X(\rho,t)$ is updated at each radial grid point and at time intervals specified by $\delta t = 10^{4}\Delta t$. The time interval is selected as $2$$\mu$s, which represents the time scale of the micro-turbulence. Therefore, the stochastic variable $X$ is uncorrelated for times larger than $\delta t$, and radial distances greater than $\delta \rho$. 

With this prescription, the vorticity Eq. \ref{1} is no longer a deterministic evolution equation and is a Langevin type stochastic differential equation. In the following we present the results of numerical simulations solving the Eqs. \ref{1} and \ref{2}.


\section{Results for passive stochastic ZFs model} \label{a}
\begin{figure}[h]
\centering
\includegraphics[width = 0.4\textwidth]{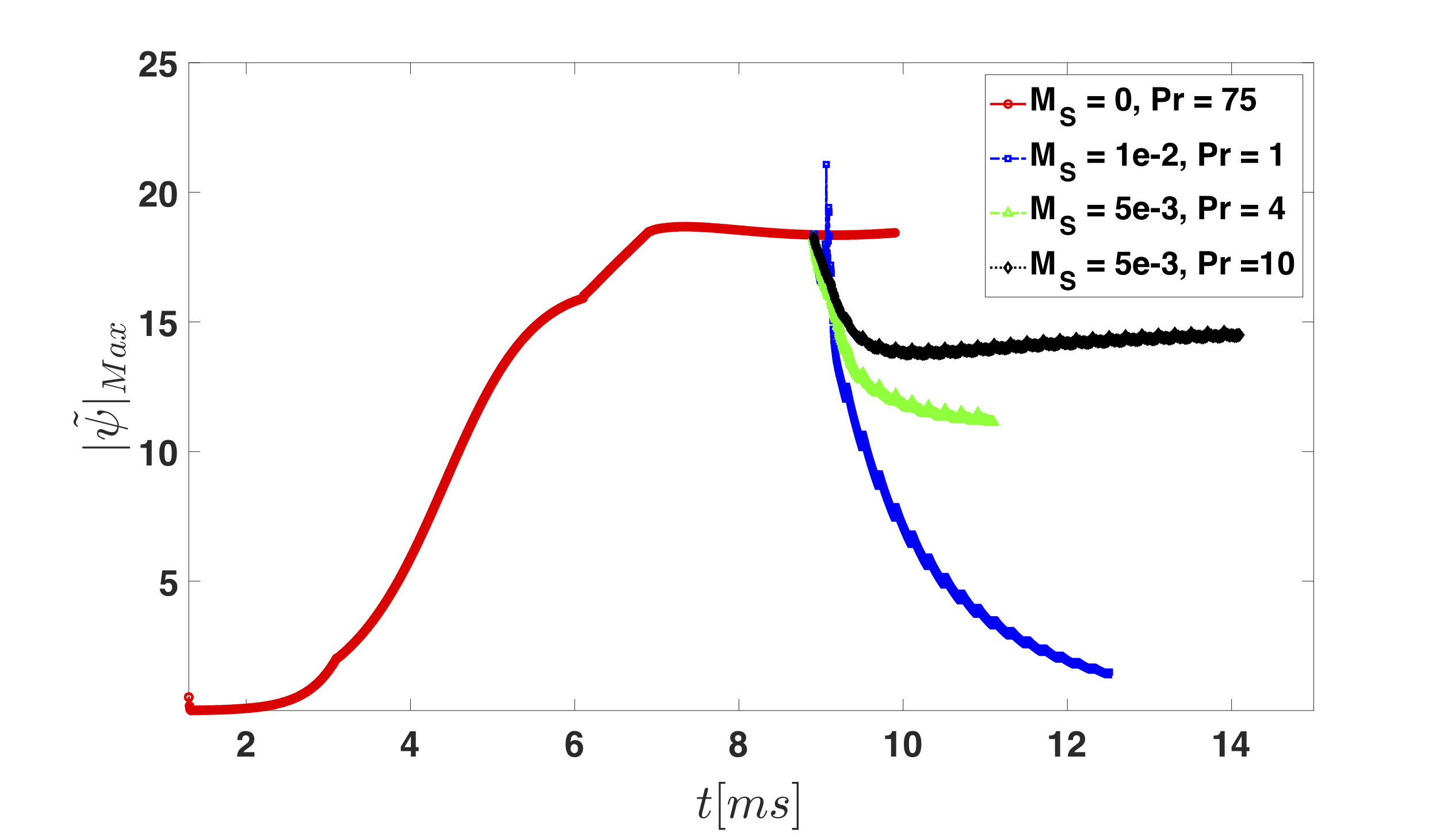}
\caption{\label{fig1} Time evolutions of the $|\tilde{\psi}|_{Max}$ for (2,1) mode for various levels of Prandtl number $Pr$, and the amplitude of the stochastic ZFs, $M_S$.}
\end{figure}

The simulations are performed by setting the initial condition, i.e. $q$-profile, to allow for the $(2,1)$ mode to be linearly unstable. In the absence of the stochastic ZFs i.e. $\delta v_S =0$, by adjusting the $Pr$ as high as $75$, a non-linearly saturated state can be reached, as seen in Fig. \ref{fig1} where the time evolution of the $|\tilde{\psi}|_{max}$ is shown (see red solid line). 

However, when the stochastic ZFs are included, the $(2,1)$ mode can be stabilised even at significantly lower Prandtl numbers. For example taking $M_S = 5\times 10^{-3}$ even at $Pr =10$ the mode reaches a saturated state at a lower amplitude, see the black dotted line in Fig. \ref{fig1}.  

The relationship between system stability through $Pr$ and $M_S$ (always keeping Lundquist number $S$ fixed) shows some interesting features. For $Pr < 2$, to stabilise the (2,1) mode the value of $M_S$ has to increase. For example, at very low $Pr = 1.25$, the stochastic ZFs Mach number, $M_S$, has to be increased to $10^{-2}$ in order to avoid the blow up of the mode, see dashed-dotted blue line in Fig. \ref{fig1}. The amplitude of the (2,1) mode in this case, is significantly reduced, and it saturates at a finite value. 

In a range of $Pr > 4$ and for the same value of $M_S$, the mode saturates at lower amplitudes for lower values of $Pr$, see for example green dashed dotted line vs black dotted line in Fig. \ref{fig1}, corresponding to $Pr=4$ and $Pr=10$, respectively. We note that the simulations show that as the $Pr$ is lowered, while keeping the $M_S$ fixed, the saturation amplitude is lower. Whilst it would be reasonable to expect that the saturation level to be higher at the lower values of viscosity (i.e. lower $Pr$), the opposite is observed.

A possible explanation for this behaviour is that at high values of $Pr$, the turbulence driven small scales will be strongly dampened by the dissipative effect of the viscosity since the damping effect of viscosity scales as $|k|^2$. Therefore the impact of small scale driven ZFs are diminished at these high values of $Pr$. At lower values of $2<Pr<4$, this damping effect is {\it reduced} and ZFs are more {\it efficient} at stabilising the (2,1) mode. 

\begin{figure}[h]
\centering
\includegraphics[width = 0.25\textwidth]{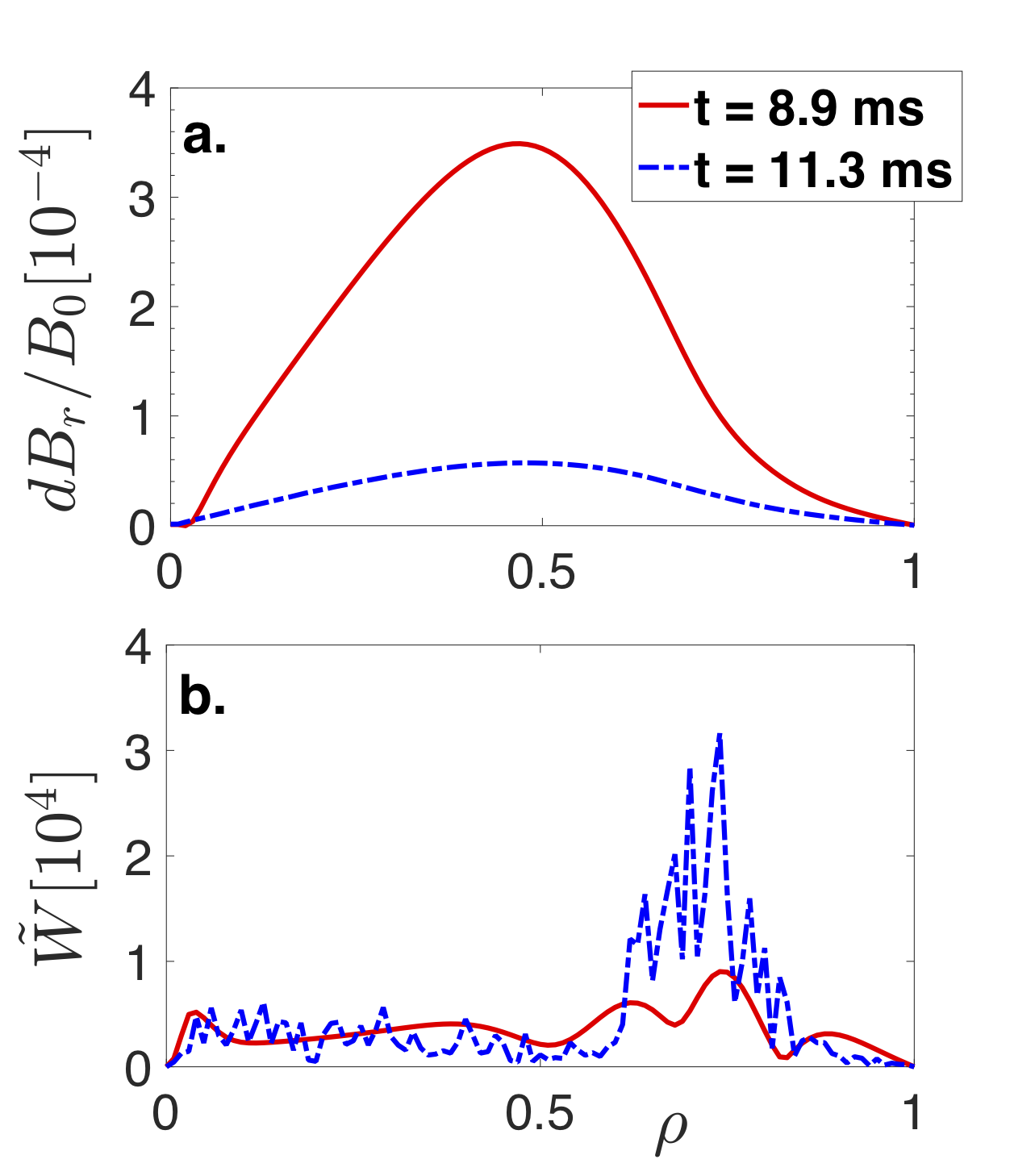}\includegraphics[width = 0.25\textwidth]{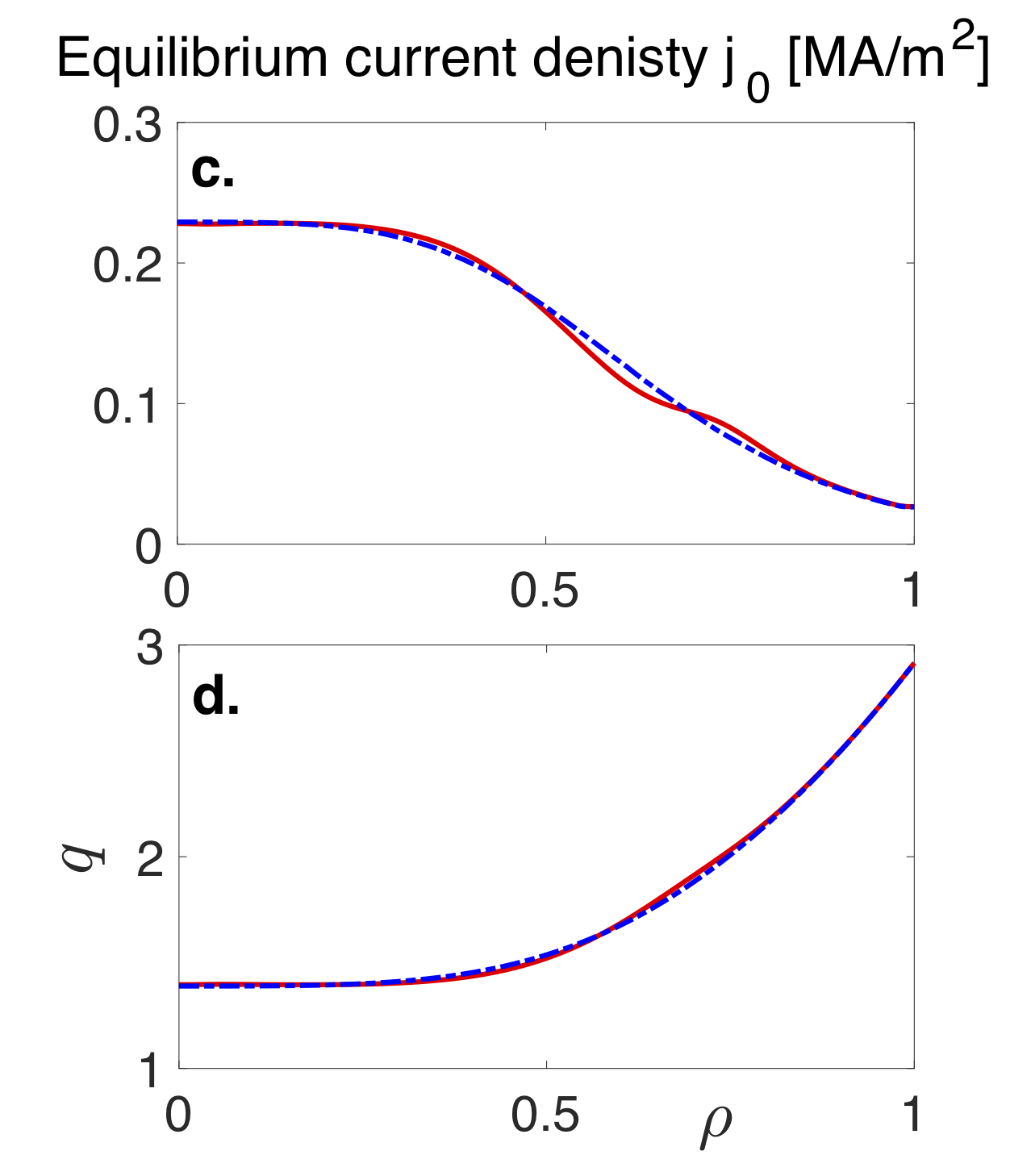} 
\caption{\label{fig2} The profiles of the $dB_r/B_0$ (a), $\tilde{W}$ (b), equilibrium current density $j_0$ (c), and safety factor $q$ (d) for the $(2,1)$ mode case, at time $t = 8.9$ms corresponding to $M_S= 0$ (red solid lines), and $t = 11.3$ms corresponding to $M_S= 10^{-2}$ (blue dotted lines).}
\end{figure}

Figures \ref{fig2} (a-d) show the corresponding profiles of the $dB_r/B_0$, $\tilde{W}$, equilibrium current density $j_0$, and safety factor $q$, for the case with $Pr =1.25$. The values are compared between $t=8.9$ms and $t=11.3$ms, without and with the stochastic ZFs effects, respectively. As can be seen here, stochastic ZFs can result in a significant stabilisation of the $(2,1)$ mode, comparing the red solid line to the blue dash-dotted line. In the pre ZFs state, $j_0$ shows local flattening around $q=2$ resonant surface, which is changed by the ZFs almost back to the linearly unstable original state. Thus, in effect, the ZFs are actually stabilising the (2,1) linearly unstable mode. The vorticity profile in this case shows peaks with higher amplitudes than in the absence of the stochastic ZFs model at $\rho \sim 0.6 -  0.8$, corresponding to the radial location of the $(2,1)$ mode (see blue dashed line in Fig. \ref{fig2} (b)). This indicates that the ZFs promote a direct cascade in the vorticity distribution of the (2,1) mode. The free energy of the equilibrium $j_0$, is transferred through the ZFs to the high $k$ vorticity of the mode, where viscosity is able to dissipate it, thus resulting in the lowering of the amplitude of the magnetic fluctuations.

\begin{figure}[h]
\centering
\includegraphics[width = 0.23\textwidth]{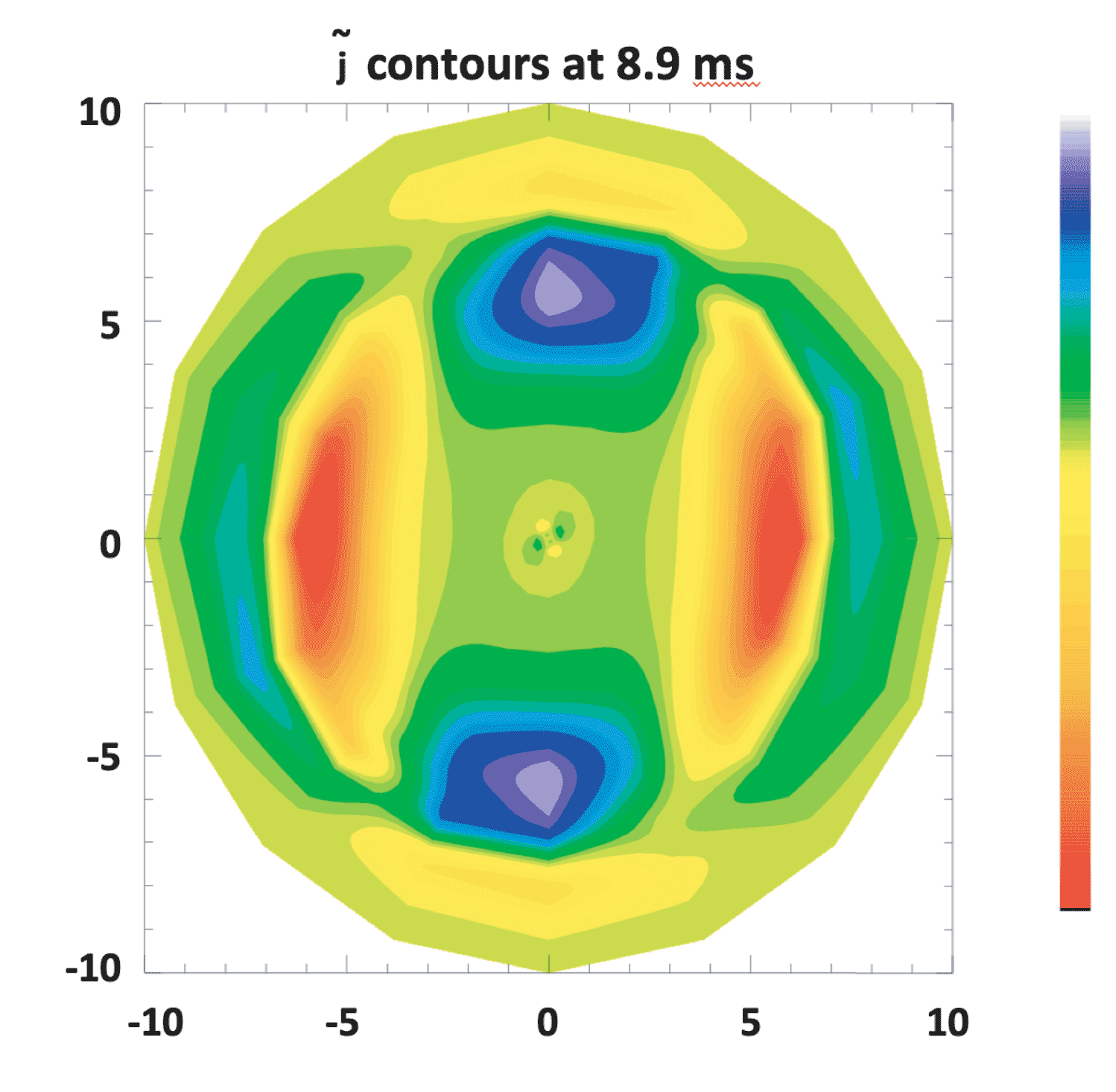}\includegraphics[width = 0.23\textwidth]{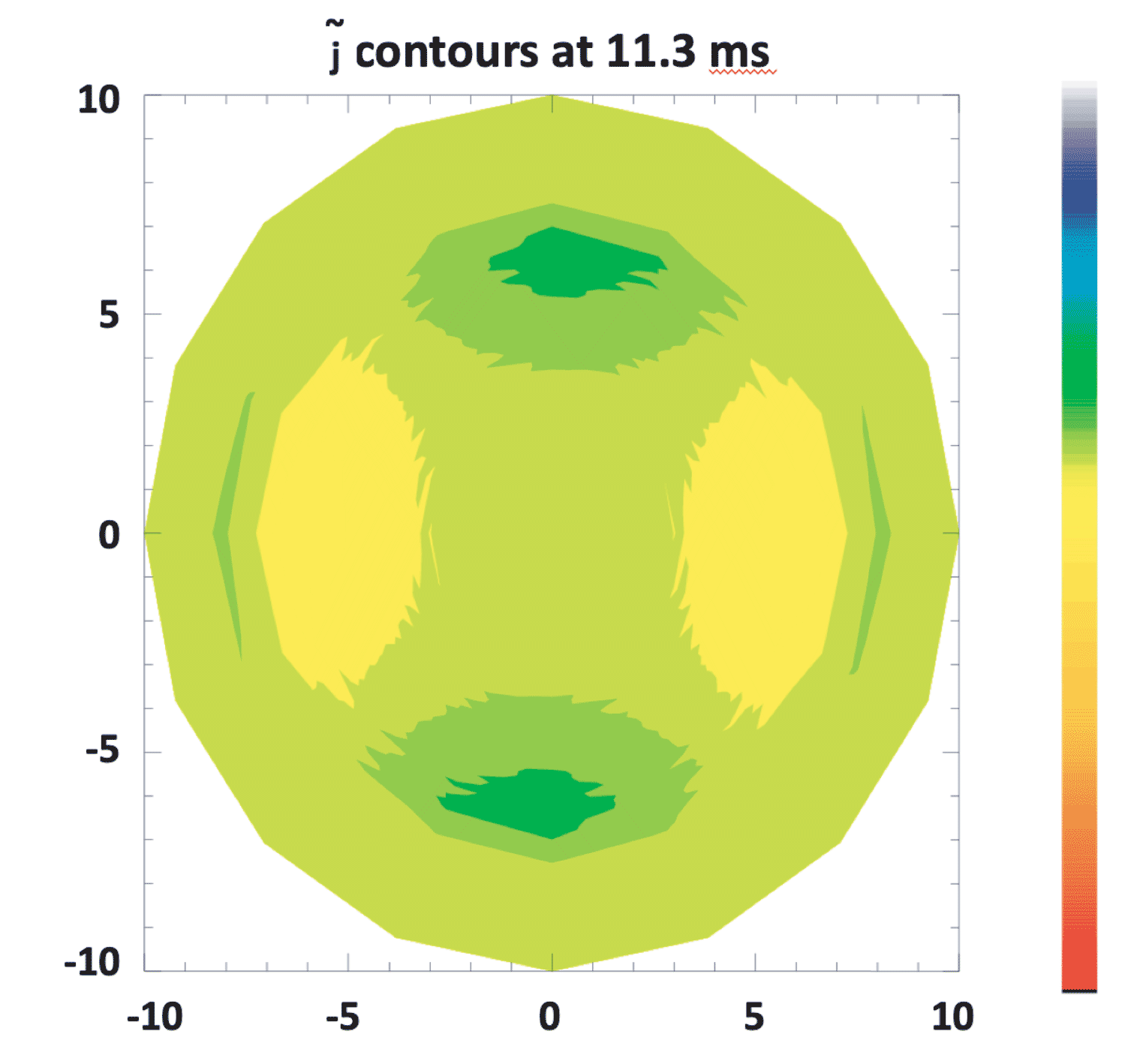}\\  
\includegraphics[width = 0.23\textwidth]{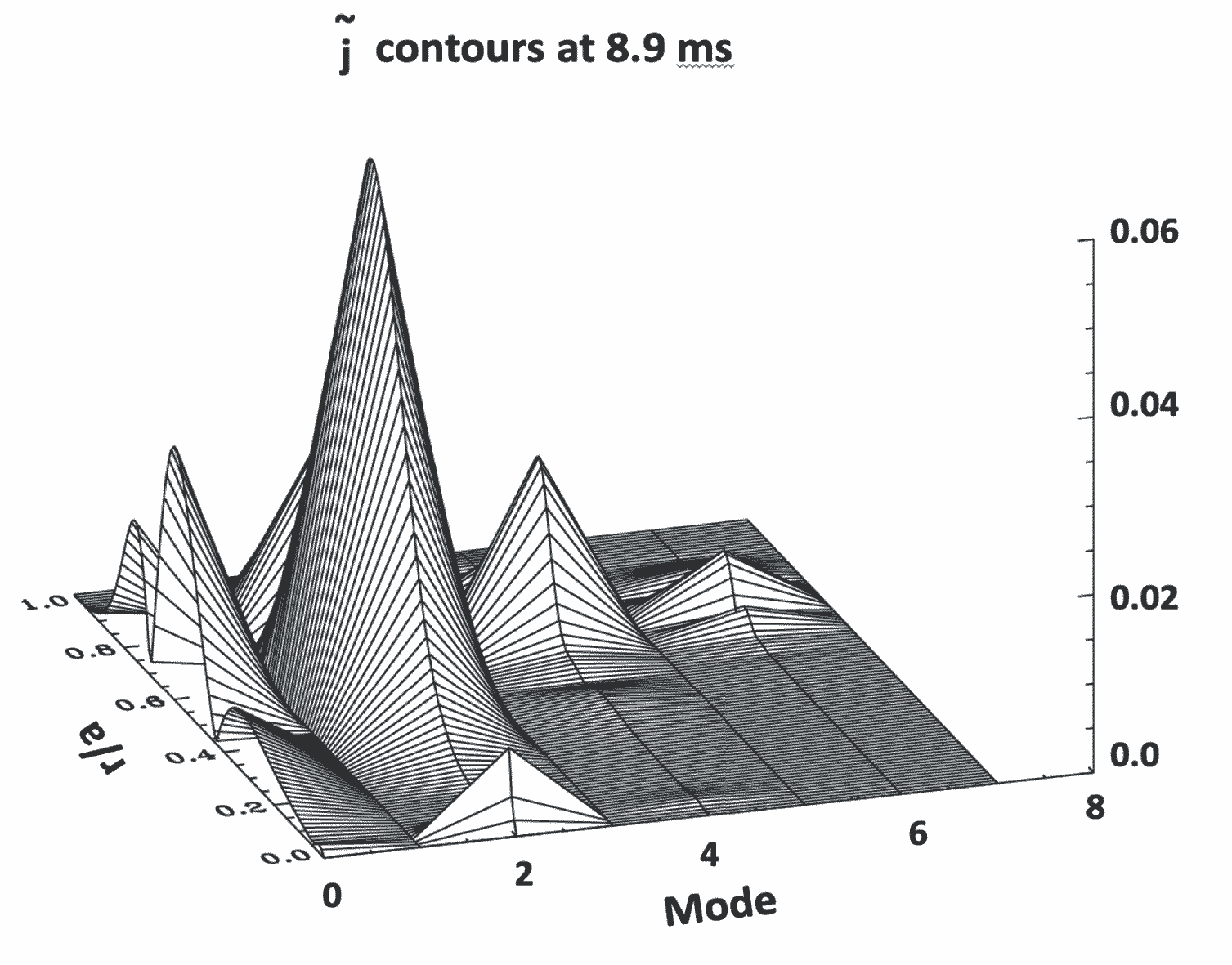}\includegraphics[width = 0.23\textwidth]{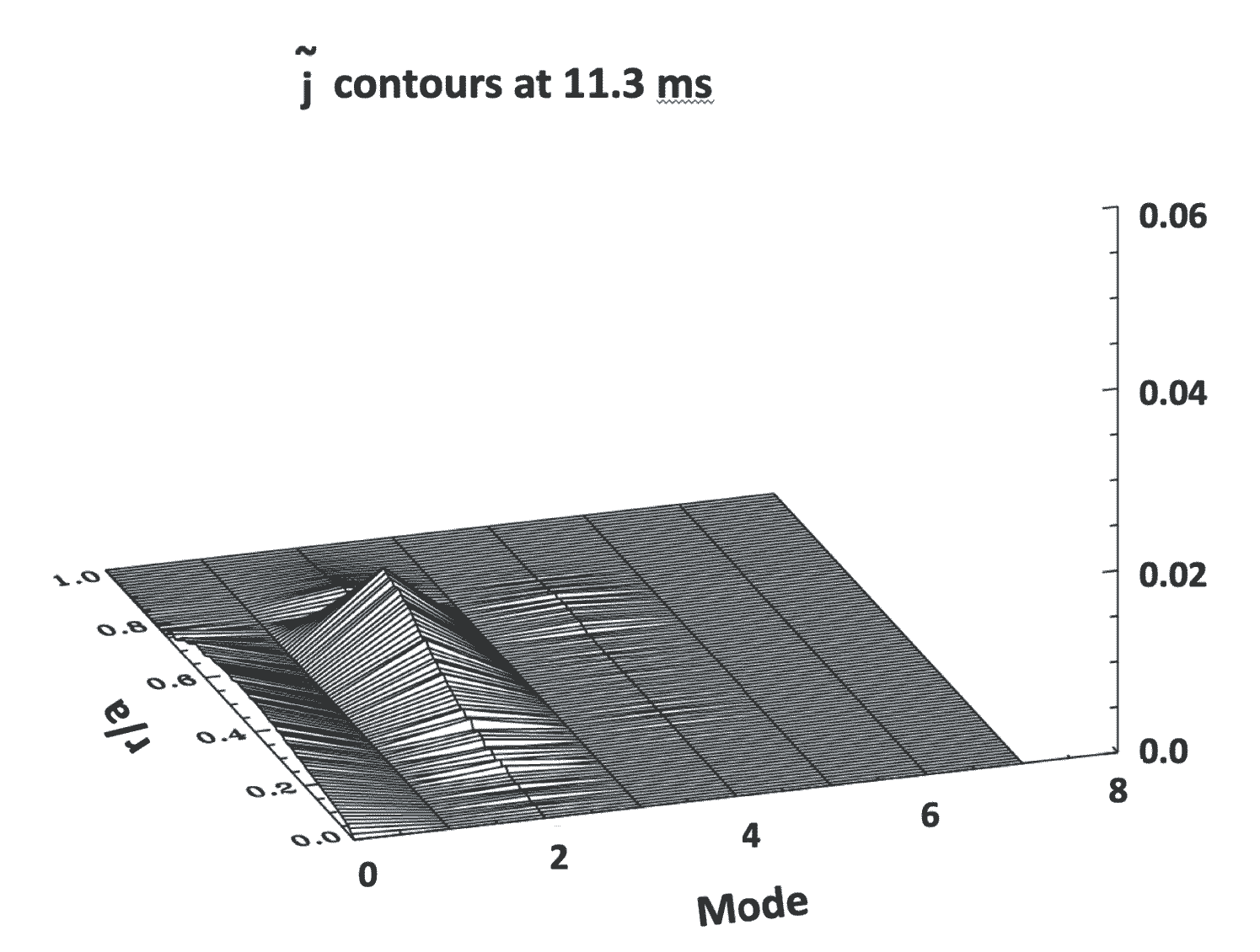} 
\caption{\label{fig3} Top: The contour plots of the current density fluctuations $\tilde{j}$ at $t=8.9$ms corresponding to the saturated level with $M_S =0, \;Pr=75$ (left), and at $t=11.3$ms with $M_S=10^{-2}, \;Pr=1.25$ (right). Bottom: The corresponding spectra of the current density fluctuations $\tilde{j}$ as functions of $\rho$ and poloidal mode number $m$, at $t=8.9$ms (left) and $t=11.3$ms (right).} 
\end{figure}

Figure \ref{fig3} illustrates the contour plots (top) and the corresponding spectra (bottom) of the current density fluctuations at the selected times. The poloidal structure of the $(2,1)$ mode remains visible at $t=11.3$ms. However, the amplitude is reduced significantly. This can also be seen in the spectra plots where a strong reduction of the current density fluctuation amplitude is obtained.

\begin{figure}[h]
\centering
\includegraphics[width = 0.4\textwidth]{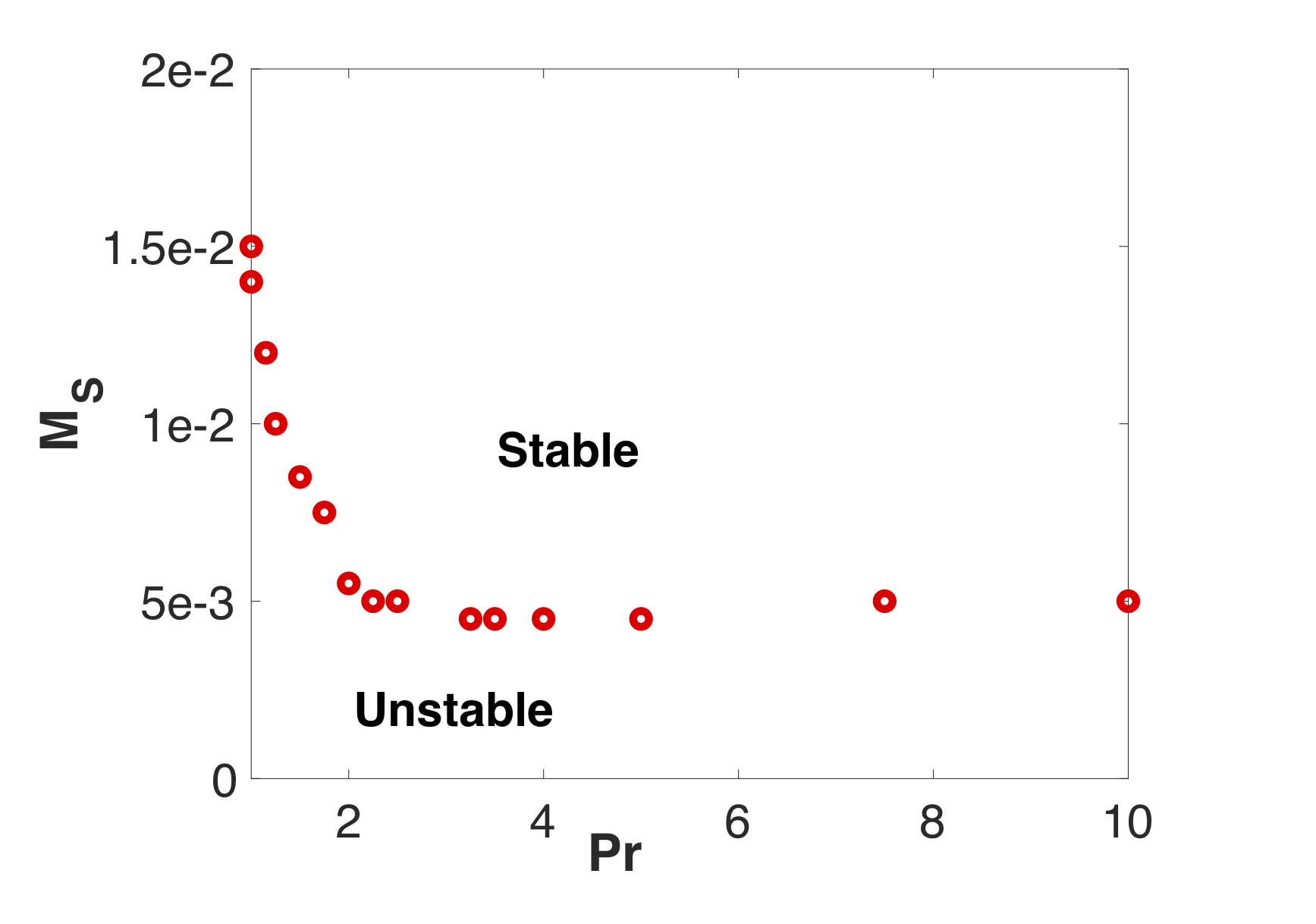}
\caption{\label{fig4} The stability boundary for (2,1) mode as functions of Prandtl number, $Pr$, and the amplitude of the stochastic ZFs, $M_S$.}
\end{figure}

To determine the stability dependence of the $(2,1)$ on the amplitude of the ZFs represented by the Mach number, $M_S$, a sensitivity scan in $(Pr,M_S)$-space was performed, while holding all other parameters fixed. The results are shown in Fig. \ref{fig4}. The stability boundary indicates that for $ 2 < Pr$ the minimum ZFs Mach number that is required to stabilise the (2,1) mode is $M_S \sim 5\times 10^{-3}$, whilst for $1 < Pr < 2$ this level has to increase, and for example for a $Pr \sim1$ it has to be as high as $M_S \sim 1.5\times 10^{-2}$. In the very low viscosity regime ($Pr<2$) mode saturation requires an inverse relationship between $Pr$ and $M_S$. 

In summary, by applying a simple passive stochastic ZFs model we were able to stabilise the $(2,1)$ tearing mode at a very low kinematic viscosity represented by the Prandtl number, $Pr$ for which the mode is linearly unstable. To stabilise the (2,1) mode for a wide range of $Pr>2$, we find that the required Mach number is around $M_S = 5\times 10^{-3}$ which is of the same order as the ZFs that is expected to be generated by the small scale turbulence.


\section{Results for semi-stochastic coupling model} \label{b}
In order to model the dynamics involved in the interaction of the visco-resistive MHD and the turbulence, in a simplified way, we have introduced a ``semi-stochastic'' coupling model, where the amplitude of the stochastic ZFs is defined as $\delta v_S = M_S v_A \langle (\tilde{j}) ^{2} + (\tilde{W})^{2}\rangle X(\rho,t) $, with $\langle \dots \rangle$ denoting average over the flux surfaces.

Here, we expect that as the amplitude of the MHD mode increases, the amplitudes of the stochastic ZFs also increase. Note that in our model, the stochasticity properties of the stochastic variable $X(\rho,t)$ are not changed. In this formulation, we attempt to model the MHD mode influence (via a direct cascade) on the interaction of the high $k$ turbulence with the mode itself. 

\begin{figure}[h]
\centering
\includegraphics[width = 0.4\textwidth]{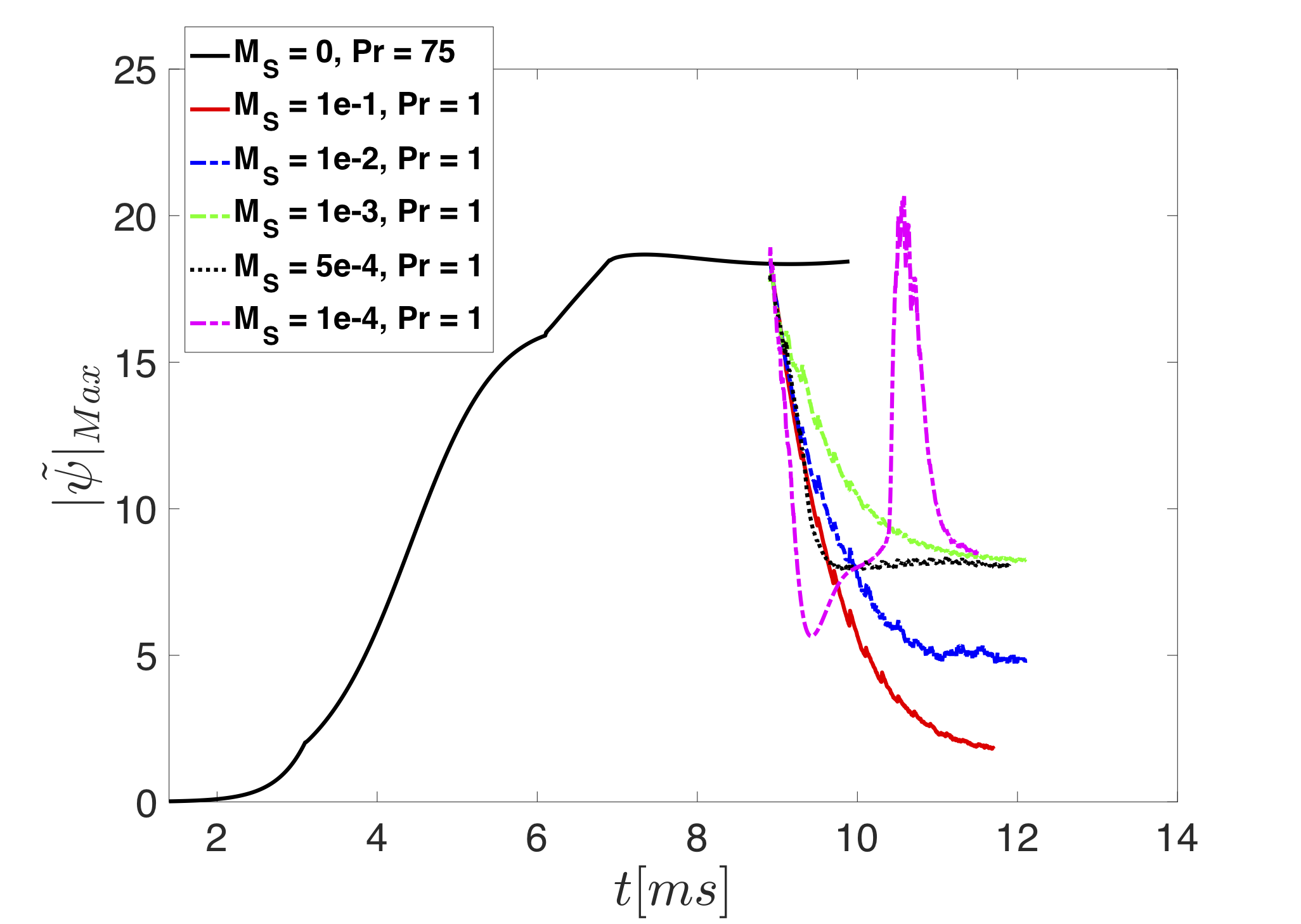} 
\caption{\label{fig5} Time evolutions of the $|\tilde{\psi}|_{Max}$ for (2,1) mode for $Pr = 75$ (black solid line) and $Pr = 1$ (all other lines), for various levels of the amplitude of the stochastic ZFs, $M_S$.}
\end{figure}

For this model, we have examined two types of MHD instabilities: i) $(2,1)$ tearing mode and ii) $(1,1)$ kink mode. This is done by setting the initial $q$ and $j_0$-profiles such that the modes are linearly unstable. The time evolution of the $|\tilde{\psi}|_{max}$ for case (i) is shown in Figs. \ref{fig5}. Here, we examine the effect on the MHD stabilisation for different $M_S$ at $Pr=1$. 

We start from the saturated phase of the $Pr=75$, $M_S=0$ (at $t = 8.9$ms) simulation, and continue while applying the semi-stochastic coupling model, and reducing the Prandtl number to $Pr = 1$ (for $t> 8.9$ms). Figure \ref{fig5} shows the results of these simulations for a wide range in $M_S = 10^{-4} - 10^{-1}$. 

Our results show that a complex dynamic interaction exists between the MHD and ZFs. Initially, as $M_S$ is decreased (from $10^{-1}$ to $10^{-3}$), the stabilisation effect of the stochastic ZFs becomes less strong, and therefore, the $(2,1)$ mode saturates at higher amplitudes; compare red solid line at $M_S = 10^{-1}$ to the green dashed line at $M_S = 10^{-3}$. However, further reduction of the $M_S$ stabilises the $(2,1)$ at a faster rate; see black dotted line in Fig. \ref{fig5} corresponding to $M_S = 5 \times 10^{-4}$. 

By decreasing the $M_S$ to $10^{-4}$ increases this initial decay rate even further. However, after reaching a threshold level, it transiently rises to a higher amplitude before eventually saturating to the same amplitude as with $M_S= 10^{-3}$; see pink dashed-dotted line in Fig. \ref{fig5} corresponding to $M_S = 10^{-4}$. A possible reason for this observation is that as the amplitude of the initial seed of ZFs becomes weaker i.e. lowering of the $M_S$, the (2,1) mode grows rapidly since at the same time we have significantly reduced the stabilisation effect of $Pr$. As the mode's growth becomes more rapid, the increase in the amplitude of the seeded ZFs grows (due to the $\langle (\tilde{j}) ^{2} + (\tilde{W})^{2}\rangle$ factor). Therefore, the stabilisation effect of the ZFs becomes stronger and stronger increasing the damping rate of the $|\tilde{\psi}|_{max}$, as observed in Fig. \ref{fig5} black dotted line compared to green dashed line. 

Consider the difference between the black dotted line ($M_S = 5\times10^{-4}$) and the pink dashed-dotted line ($M_S = 10^{-4}$), in Fig. \ref{fig5}. The rapid transient growth of the mode amplitude due to the significantly lower $M_S$ results in a very large rise of the non-linear factor $\langle (\tilde{j}) ^{2} + (\tilde{W})^{2}\rangle$. As expected, the mode drives the ZFs strongly leading to even faster damping. This results in the pink dashed-dotted line reaching a minimum at about $t\sim 9.2$ms well bellow the black dotted line, in Fig. \ref{fig5}. On the other hand, if the mode decays too rapidly to very low levels, the ZFs amplitude decays with it and so does its stabilisation impact on the (2,1) mode. As soon as the ZFs stability effect disappears, the non-linear growth of the mode rapidly results in an overshoot (see pink dashed-dotted line in Fig. \ref{fig5} at $t=10.4$ms where the mode has reached reaching its maximum at $t\sim 10.4$ms well above the saturation level of the black dotted line). At this point, the ZFs become large enough to suppress the non-linear growth and stabilise the system to reach a new saturated state which is essentially similar to the previous cases. For a range of initial ZFs seeds with $M_S = 10^{-4}-10^{-3}$, we observe a non-linear self-organziation interplay between ZFs, $(2,1)$ and its related low (m,n) spectrum that saturates to a similar final state.     
 
\begin{figure}[h]
\centering
\includegraphics[width = 0.25\textwidth]{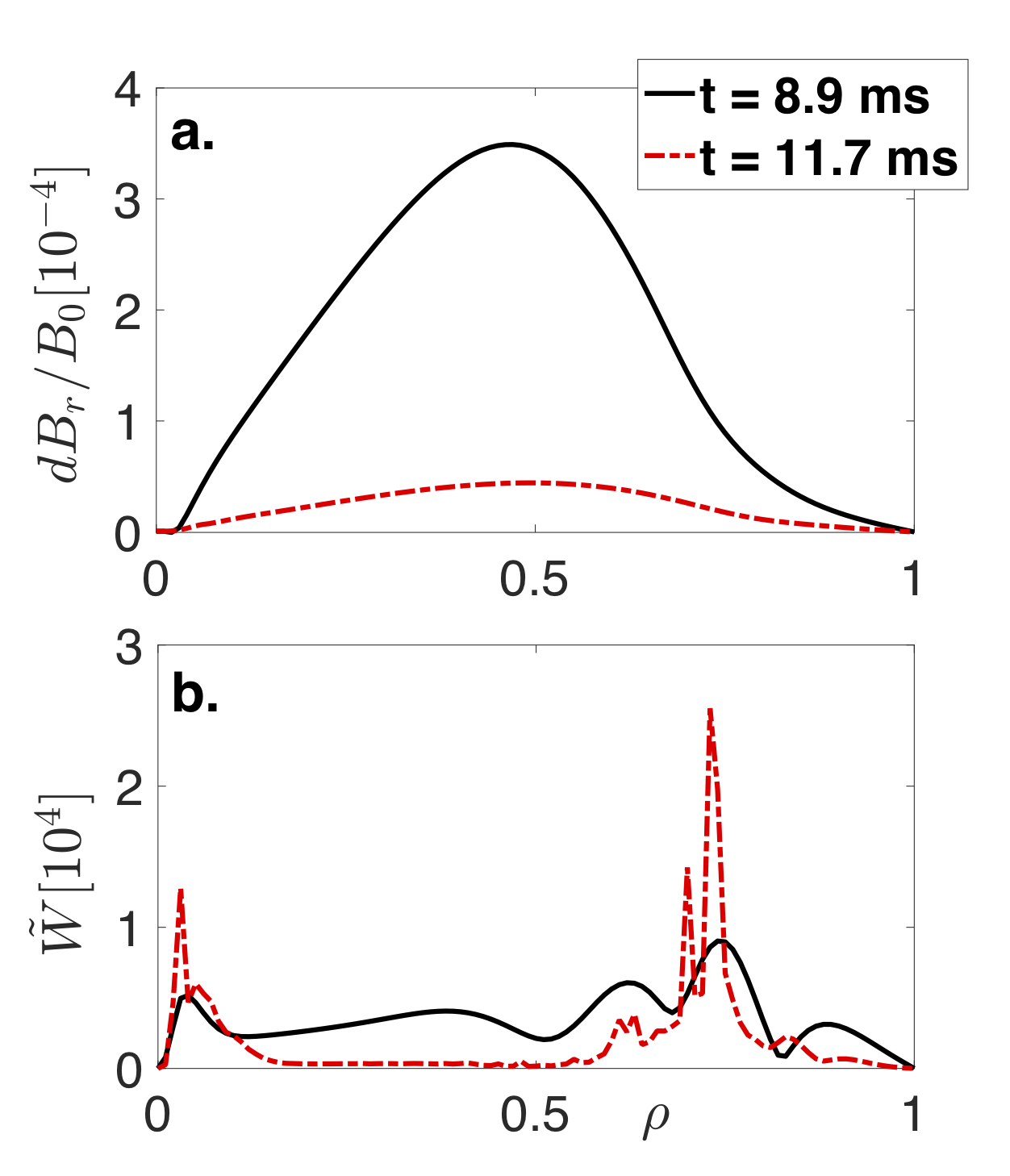}\includegraphics[width = 0.25\textwidth]{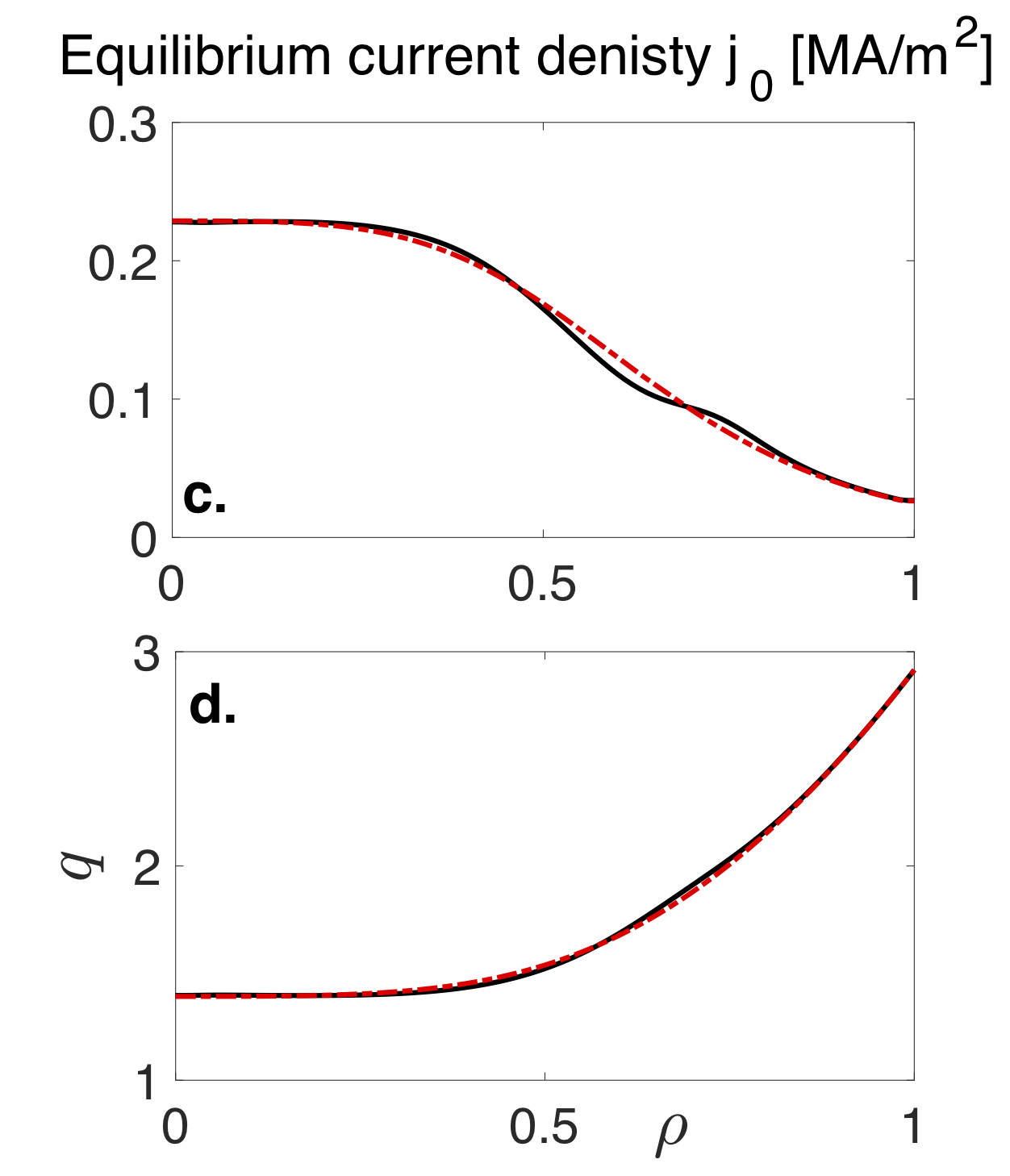}
\caption{\label{fig6} The profiles of the $dB_r/B_0$ (a), $\tilde{W}$ (b), equilibrium current density $j_0$ (c), and safety factor $q$ (d) for the $(2,1)$ mode case, at time $t = 8.9ms$ corresponding to $M_S = 0$ (red solid lines), and $t = 11.7ms$ corresponding to $M_S = 10^{-1}$ (blue dotted lines).}
\end{figure}

Figures \ref{fig6} (a-d) show a comparison of the profiles of the $dB_r/B_0$, $\tilde{W}$, equilibrium current density $j_0$, and safety factor $q$ between $t=8.9$ms and $t=11.7$ms, without and with the stochastic ZFs, respectively. 
The equilibrium current profile $j_0$ in the final saturated state, is almost back to its original linearly unstable profile, in the case where $M_S = 10^{-1}$ (see Fig. \ref{fig6} (c)). However, the saturated mode amplitude is low $dB_r/B_0 \sim 10^{-4}$, thus, the overall ZFs Mach number is of the order of $10^{-3}$.

\begin{figure}[h]
\centering
\includegraphics[width = 0.23\textwidth]{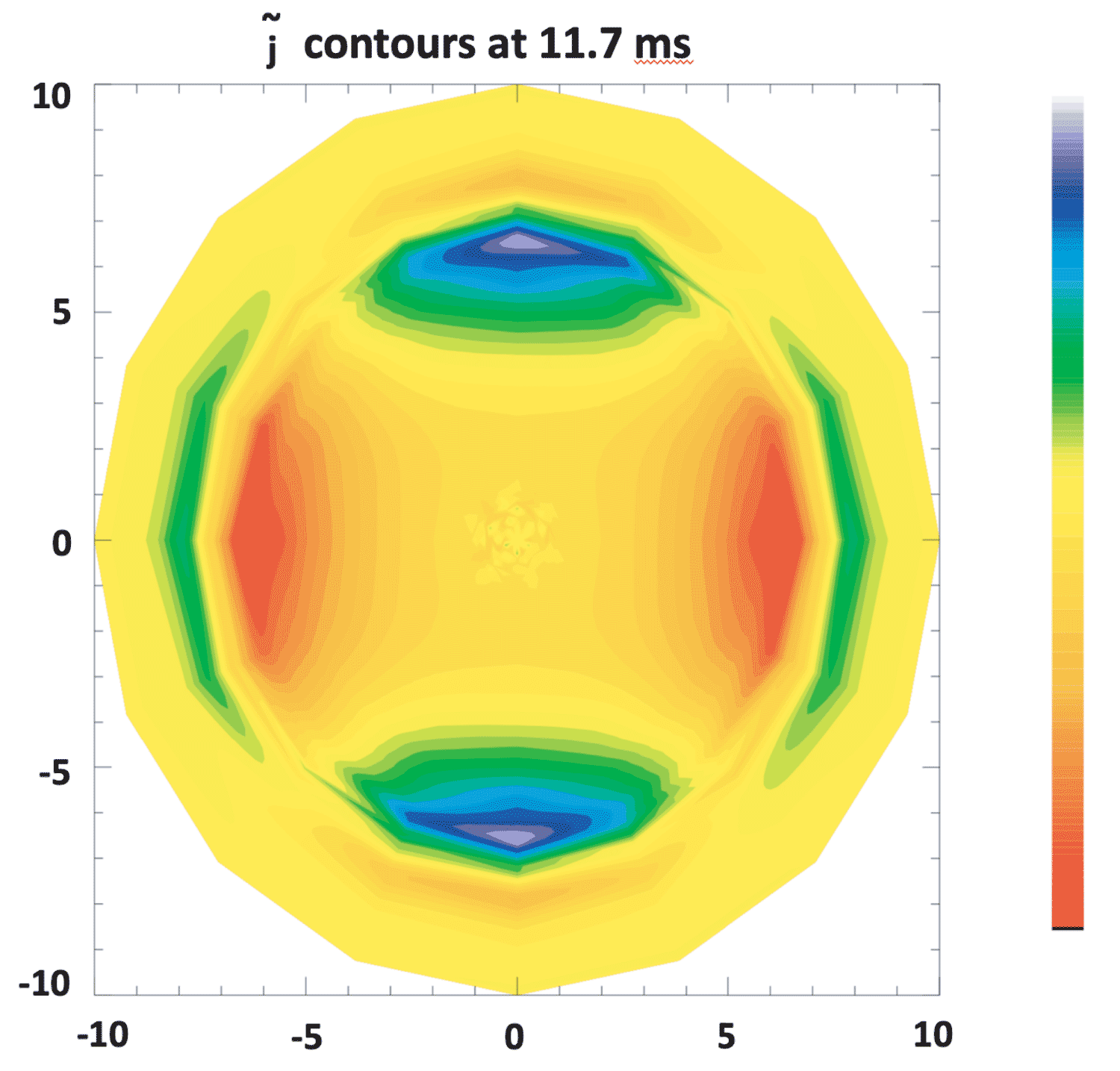}\includegraphics[width = 0.23\textwidth]{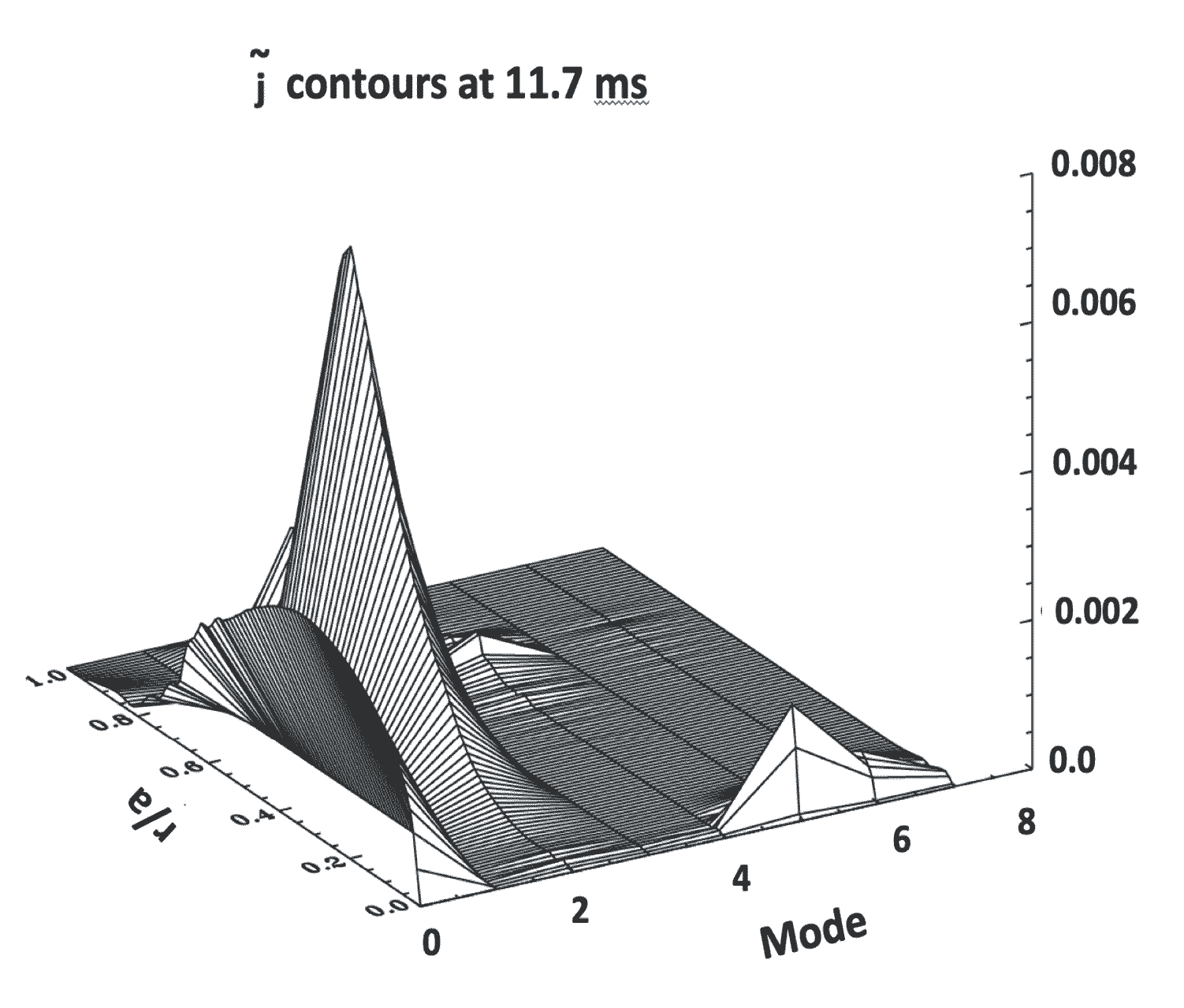}
\caption{\label{fig7} Left: The contour plots of the current density fluctuations $\tilde{j}$ at $t=11.7$ms corresponding to the saturated level with $M_S =10^{-1}, \;Pr=1$. Right: The corresponding spectra of the current density fluctuations $\tilde{j}$ as functions of $\rho$ and poloidal mode number $m$.} 
\end{figure}

The vorticity profile in this case shows two distinct peaks with significantly higher amplitudes than in the absence of the stochastic ZFs at $\rho \sim 0.65$ and $\rho \sim 0.75$ (see blue dashed line in Fig. \ref{fig6} (b)). The impact of this strong shear in the vorticity is observed on the sheared poloidal structure of the current density fluctuations with two peaks developed at $m=2$ surface, see the contour plots shown in Fig. \ref{fig7}(left) and the corresponding spectra plots in Fig. \ref{fig7}(right). Furthermore, a low amplitude $m=5, \; n=4$ mode is found to be active in the plasma core. 

\begin{figure}[h]
\centering
\includegraphics[width = 0.4\textwidth]{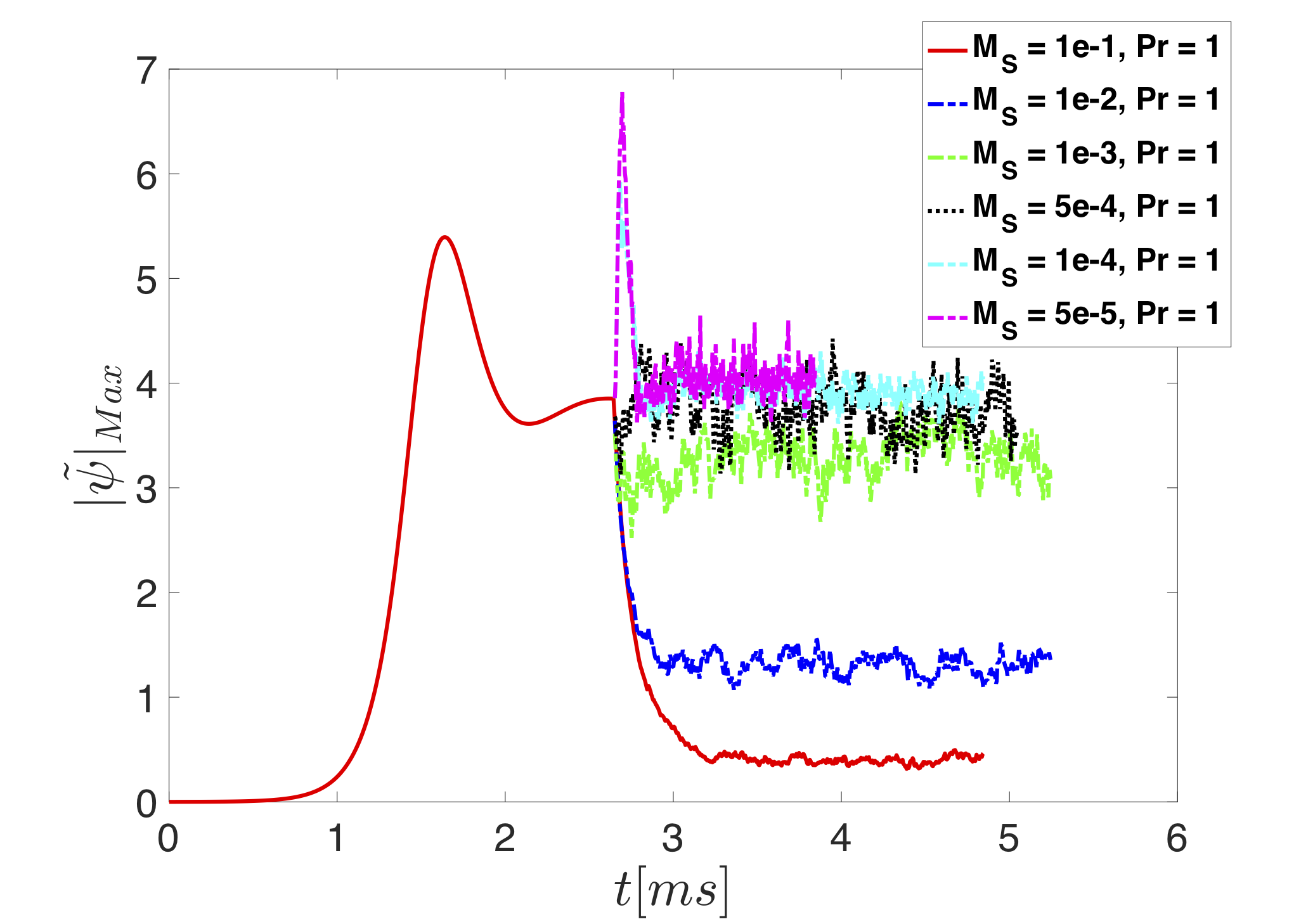}
\caption{\label{fig8} Time evolutions of the $|\tilde{\psi}|_{Max}$ for (1,1) mode for $Pr = 30$ (black solid line) and $Pr = 1$ (all other lines) for various levels the Mach number of the stochastic ZFs, $M_S$.}
\end{figure}

\begin{figure}[h]
\centering
\includegraphics[width = 0.25\textwidth]{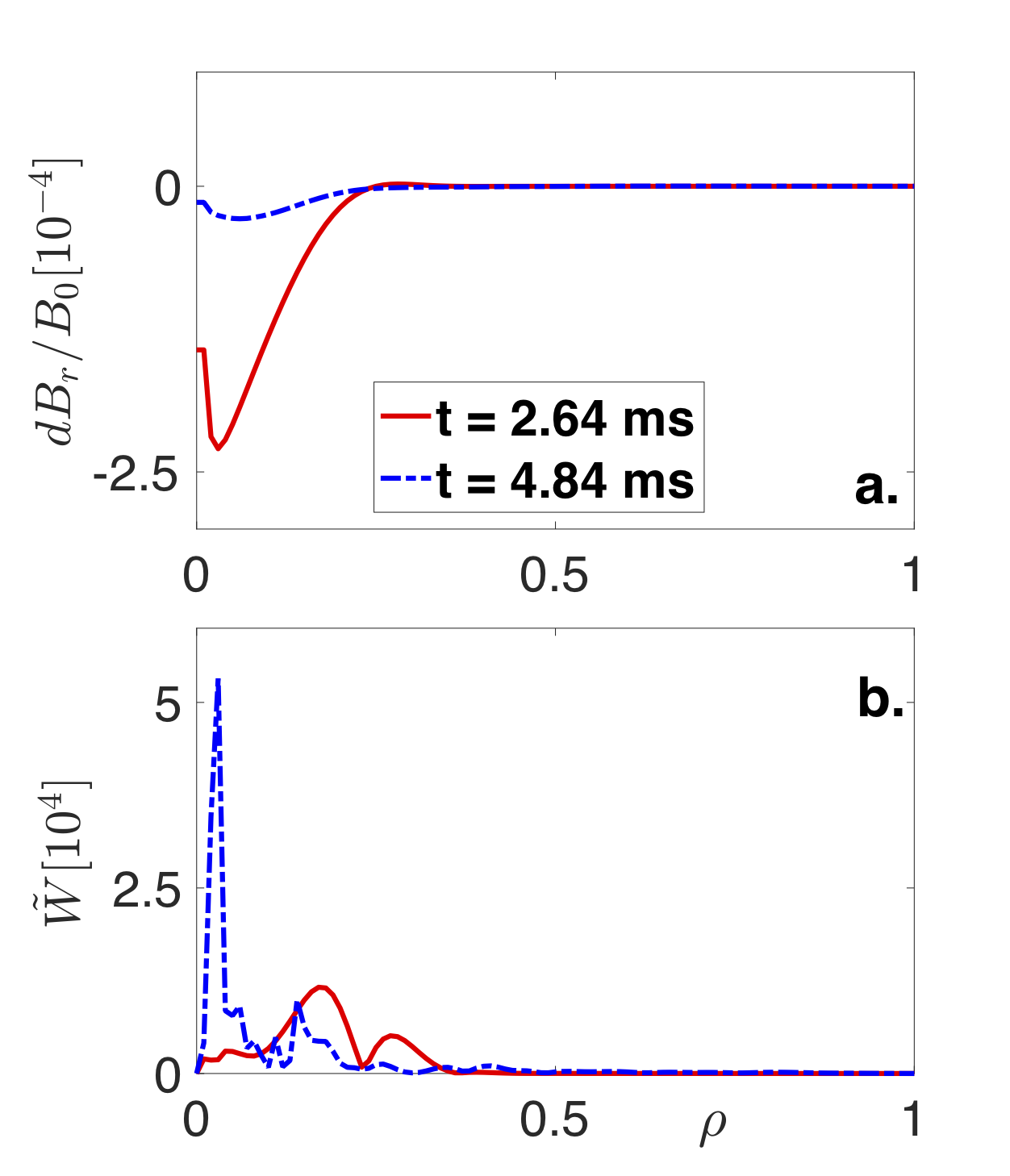}\includegraphics[width = 0.25\textwidth]{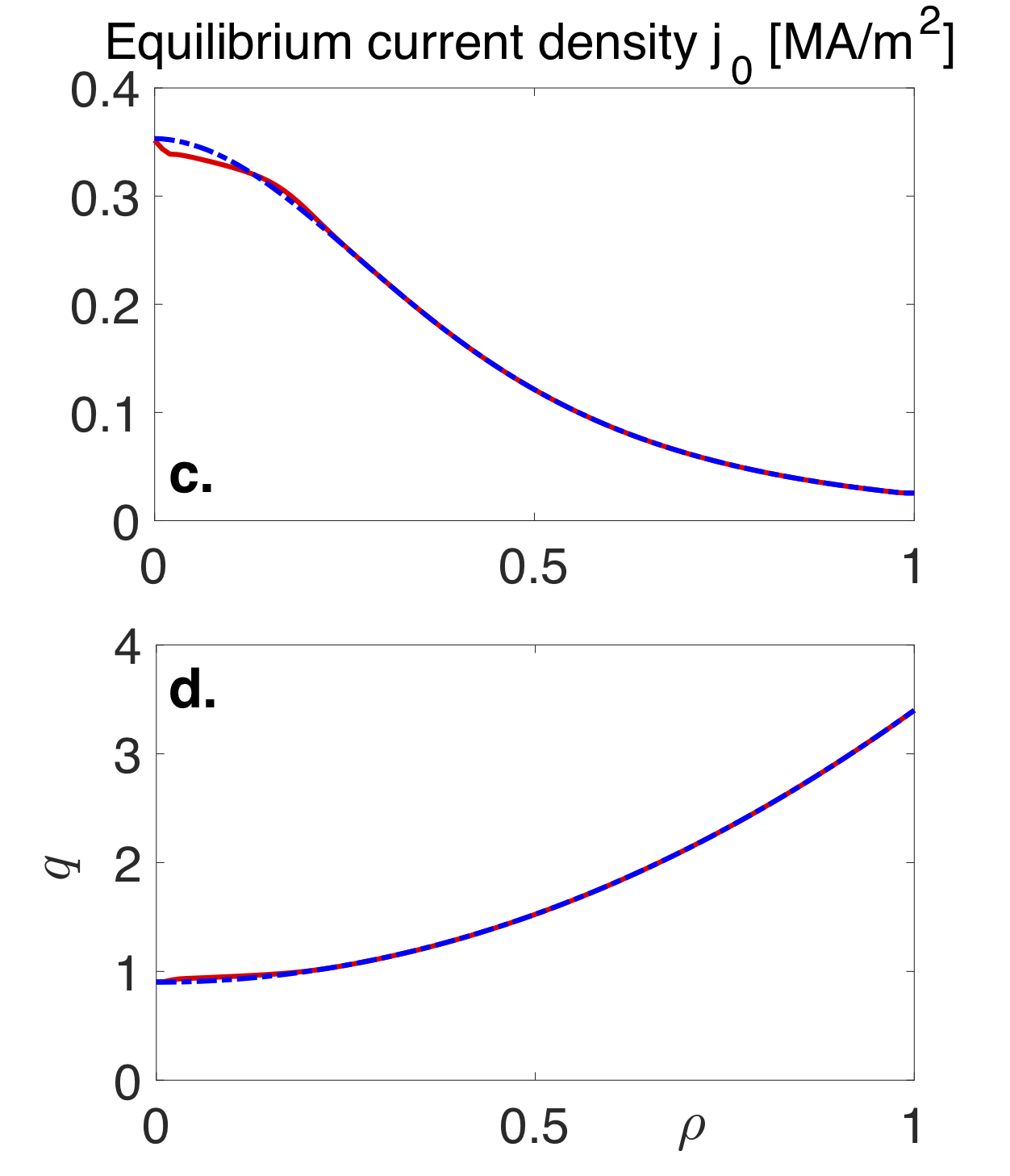} 
\caption{\label{fig9} The profiles of the $dB_r/B_0$ (a), $\tilde{W}$ (b), equilibrium current density $j_0$ (c), and safety factor $q$ (d) for the $(1,1)$ mode case, at time $t = 2.64$ms corresponding to $M_S= 0$ (red solid lines), and $t = 4.64$ms corresponding to $M_S = 10^{-1}$ (blue dotted lines).}
\end{figure}

In case ii) for the $(1,1)$ mode we start with the Prandtl number of $Pr = 30$ (as was reported in Ref. \cite{Mendonca2018}), and after reaching a non-linear saturated state at time $t=2.64$ms, we apply the stochastic ZFs model as described above whilst reducing the Prandtl number to $Pr=1$. As can be seen in Fig. \ref{fig8}, for very small values of the $M_S = 5\times10^{-5}$ (pink dashed-dotted line), initially the $(1,1)$ mode grows strongly due to the lower Prandtl number, however as its amplitude increases, so does the amplitude of the stochastic ZFs, which eventually leads to stabilisation and saturation of the mode. We note that during this mode evolution, the equilibrium state also changes ($dj_0/dr$) due to the non-linear effects of the mode on itself, the so-called profile-mode interaction \cite{Mendonca2018}. After that the mode fluctuates around its original saturated level in the absence of the stochastic ZFs with $Pr=30$. By increasing $M_S$, a new saturated state can be reached but at a significantly lower amplitude (see for example, red solid line in Fig. \ref{fig8} for $M_S= 10^{-1}$). This dynamic behaviour is closely related to the dynamics that we described in the (2,1)
 case where the equilibrium current profile $j_0$ in the final saturated state, is almost back to its original linearly unstable profile (see Fig. \ref{fig9} (c)). However, the saturated mode amplitude is low $dB_r/B_0 \sim 10^{-5}$, thus, the overall ZFs Mach number is agin of the order of $10^{-3}$.
 
\begin{figure}[h]
\centering
\includegraphics[width = 0.23\textwidth]{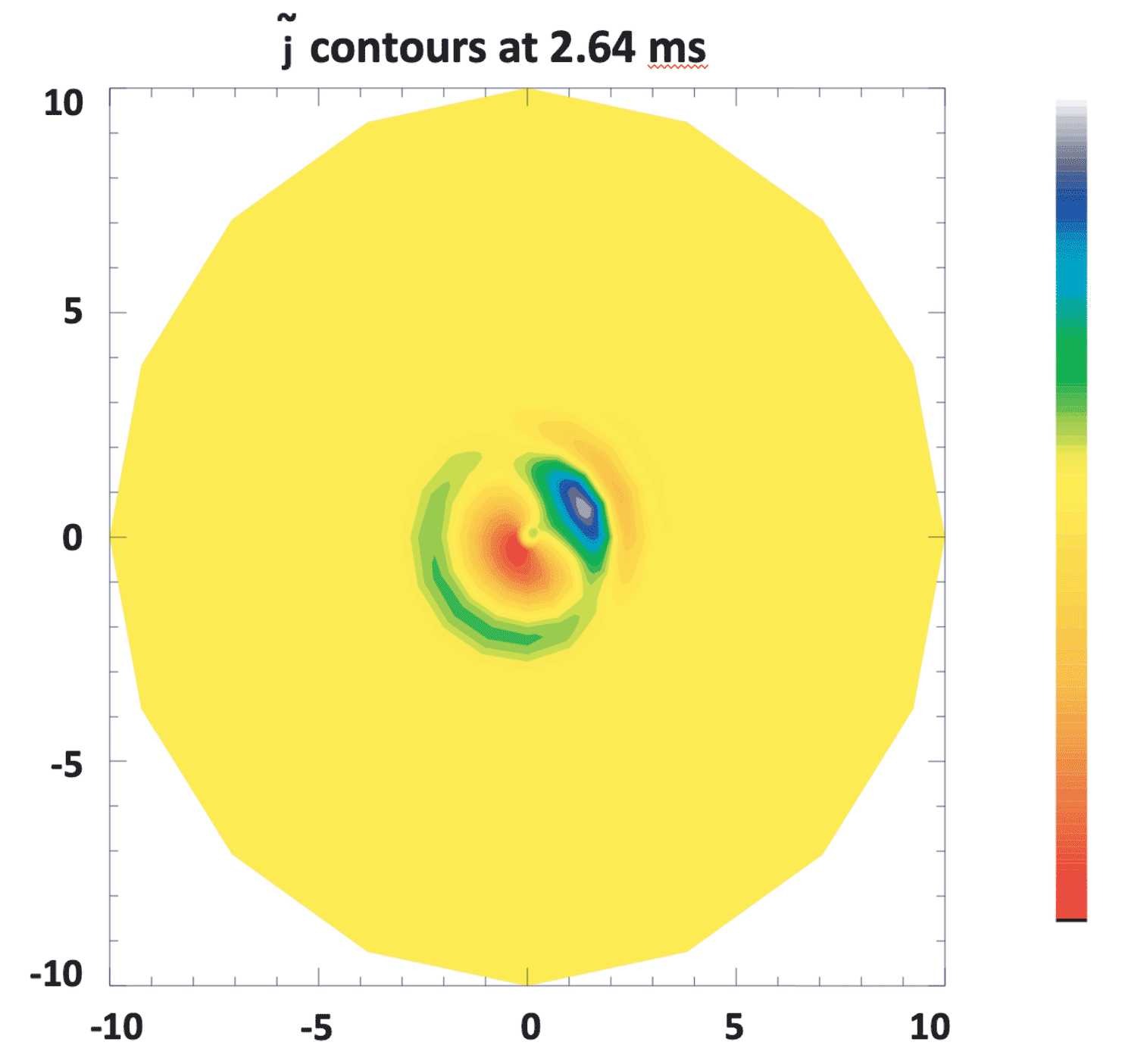}\includegraphics[width = 0.23\textwidth]{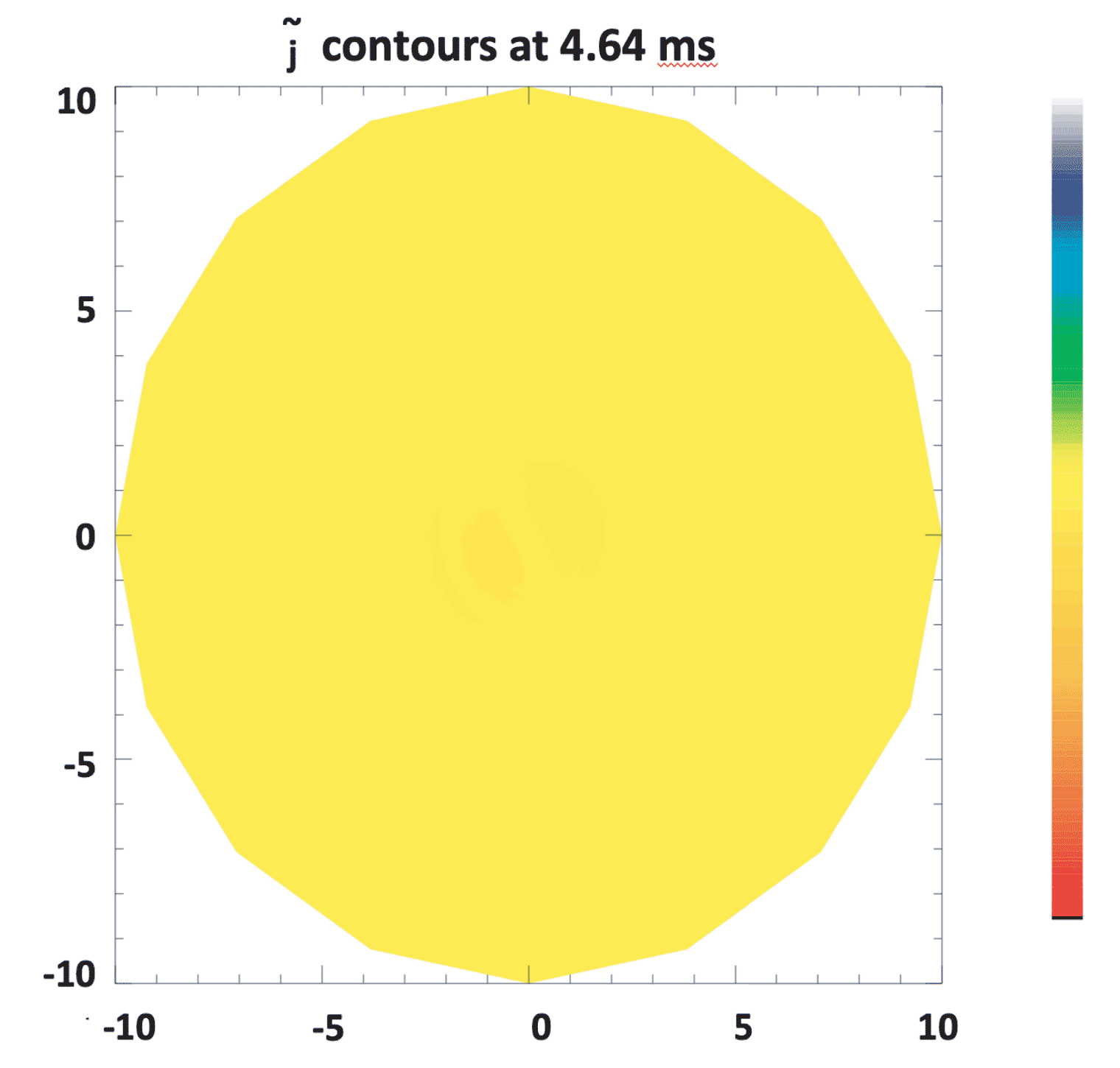}\\ 
\includegraphics[width = 0.23\textwidth]{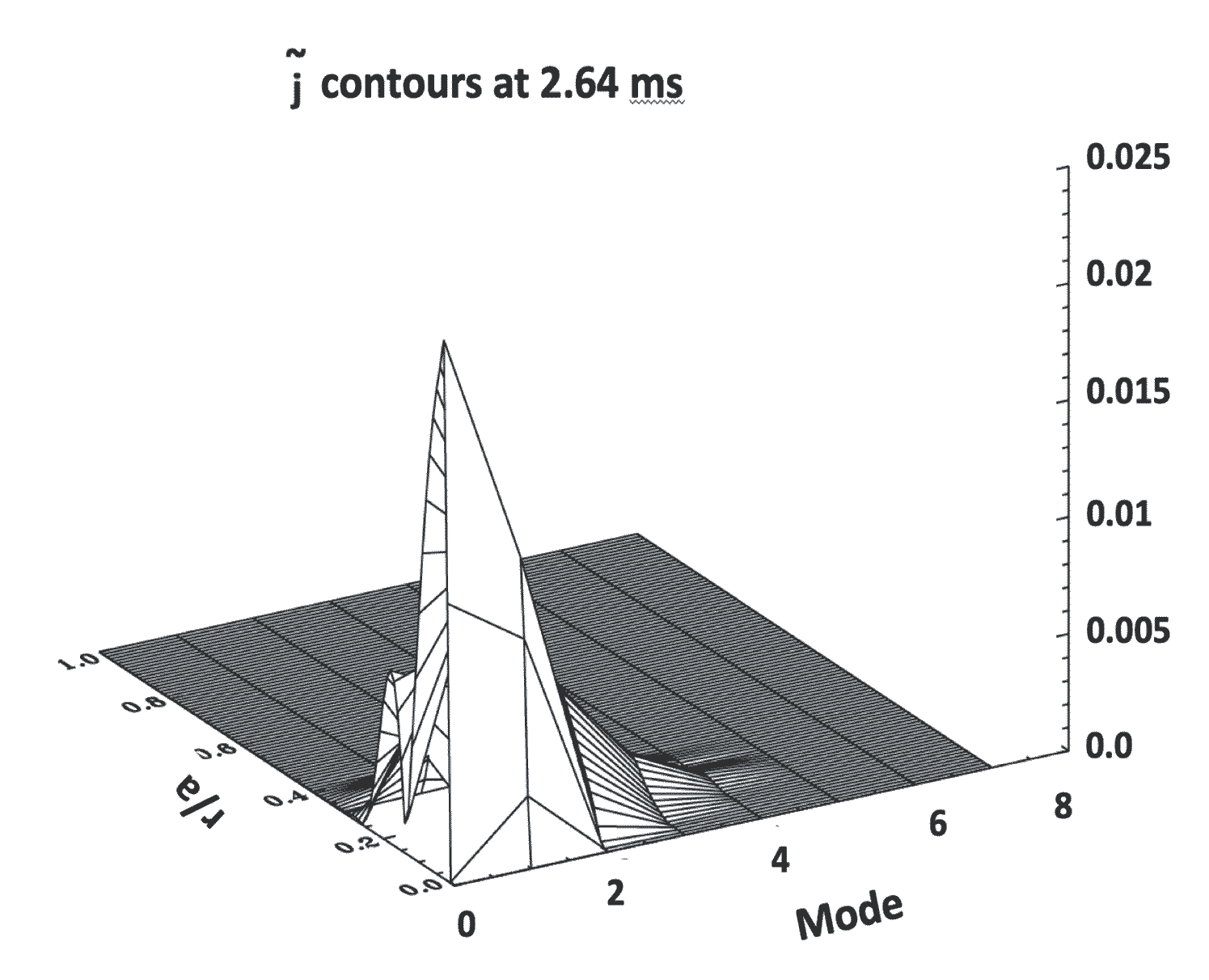}\includegraphics[width = 0.23\textwidth]{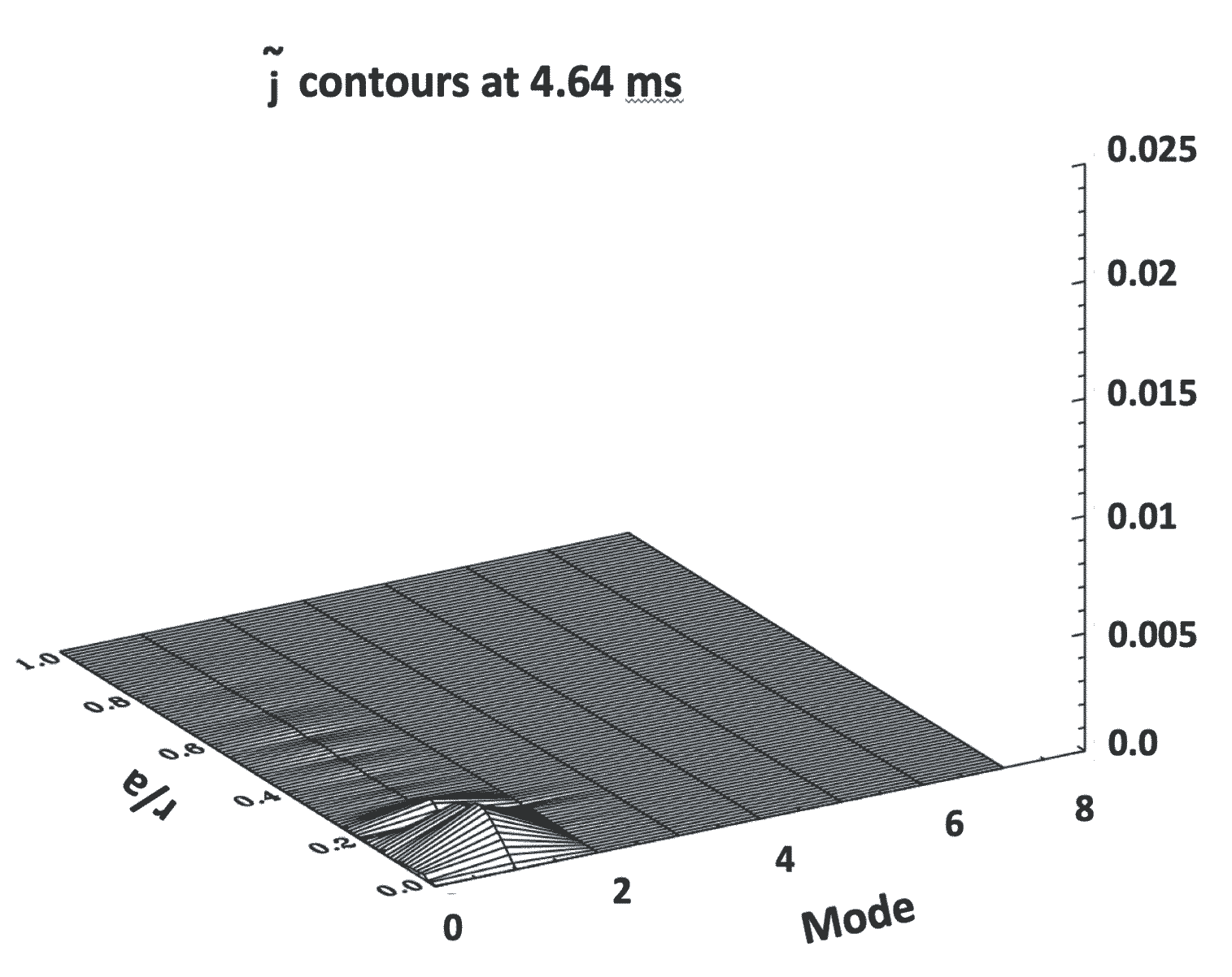}\\ 
\caption{\label{fig10} Top: The contour plots of the current density fluctuations $\tilde{j}$ at $t=2.64$ms corresponding to the saturated level with $M_S =0, \;Pr=30$ (left), and at $t=4.64$ms with $M_S =10^{-1}, \;Pr=1$ (right). Bottom: The corresponding spectra of the current density fluctuations $\tilde{j}$ as functions of $\rho$ and poloidal mode number $m$, at $t=2.64$ms (left) and $t=4.64$ms (right).} 
\end{figure}

Figures \ref{fig9} (a-d) show a comparison of the profiles of the $dB_r/B_0$, $\tilde{W}$, equilibrium current density $j_0$, and the safety factor $q$ at $t=2.64$ms and $t=4.64$ms, without and with the stochastic ZFs effects at $M_S = 10^{-1}$, respectively. As can be seen here, addition of the stochastic ZFs results in a significant stabilisation of the $(1,1)$ mode, comparing the red solid line to the blue dash-dotted line in Fig. \ref{fig9} (a). The distortion of the equilibrium current density due to the $(1,1)$ mode is strongly reduced as the model is applied. The vorticity profile in this case shows a more localised peak with larger amplitude at $\rho \sim 0.15$ (see blue dashed line in Fig. \ref{fig9} (b)). The impact of this strong shear in the vorticity is observed on the poloidal structure of the current density fluctuations at $m=1$ surface, see the contour plots shown in Fig. \ref{fig10}(Top) and the corresponding spectra plots in Fig. \ref{fig10}(Bottom).

In summary, by applying a semi-stochastic ZFs model we were able to stabilise the $(2,1)$ tearing mode and the $(1,1)$ kink mode at a very low kinematic viscosity represented by the Prandtl number, $Pr$ for which the modes are linearly unstable. The main difference here with that of the simple stochastic model discussed in the previous section is that, the dynamic stabilisation is more effective even at lower $M_S$ and after an initial oscillation the saturation levels for different values of $M_S$ are similar due to the interaction between MHD and turbinate.

\section{Conclusion} 
A model is developed to simulate micro-scale turbulence driven ZFs, and their impact on the MHD tearing and kink modes is examined. The model is based on a stochastic representation of the micro-scale ZFs with a given Mach number, $M_S$. Two approaches were considered i) passive stochastic model where the ZF's amplitudes are independent of the MHD mode amplitude, and ii) the semi-stochastic model where the amplitudes of the ZFs have a dependence on the amplitude of the MHD mode itself. We have purposely kept the resistive diffusivity ($c^2\eta/4\pi$), and the kinematic viscosity ($\nu$) as given functions of $\rho$, and during the time evolution of the system, in order to explore the advective effects of the stochastic ZFs on the vorticity equation. We recognise that in principle, the small-scale turbulence can also modify both of these transport properties for example as it is done in previous two-fluid simulations with CUTIE (Thyagaraja et. al. \cite{Thyagaraja2010}). In neutral fluid dynamics, it has long been recognised (since the time of Prandtl), that turbulence modifies momentum and energy transport properties of the fluid significantly above their Laminar/molecular values.  

The results for the approach i) show that the stochastic ZFs can significantly stabilise the (2,1) mode even at very low kinematic viscosity, $Pr$, where the mode is linearly unstable. For very low levels of $Pr < 2$ the stochastic ZFs amplitude, described by its Mach number $M_S$ has to increase, in order for stabilisation to be effective. However, for $Pr>2$, and at similar values of $M_S$, we find that lower values of $Pr$ lead to lower saturation levels of the MHD mode. This is understood to be the result of stronger dissipation of the ZFs at higher $Pr$, which reduces their stabilisation effect on the MHD mode.

In the approach ii) we observe a more complex interaction between the ZFs and the MHD modes (both (2,1) and (1,1) modes), where by decreasing the amplitude of the initial seeded ZFs, i.e. $M_S$, the damping rate of the MHD modes actually increases even at linearly unstable $Pr$ levels. This dynamic behaviour is understood to be a result of a predator-prey type of dynamical stabilisation where the initial seeded ZFs are too small to damp the MHD mode. However since the stabilisation effect of the viscosity ($Pr$) has also diminished in these cases, the mode amplitude shoots up. As the amplitude of ZFs are dependent on the energy of the mode, they will grow very fast and back-react on the mode leading to a faster decay rate. For $M_S$ values below a threshold, the fast decay of the MHD also kills the ZFs, and hence, it starts to grow again. As the ZFs grow with the mode, at some point they reach a level that can stabilise the MHD mode once more. In our examples, we have not observed any repetition of this cycle, and after this initial interaction between the ZFs and MHD mode, they saturate to an invariant final state (for $10^{-5} < M_S  < 10^{-3}$). The equilibrium current density profile in this final state is close to the initial linearly unstable profile. We note however, that the possibility of a ``Hopf bifurcation" with periodic waxing and wining of the mode and turbulence is not ruled out and may occur for different choices of parameters.

Our results, therefore indicate a possible mechanism for stabilisation of the MHD modes via small-scale perturbations in poloidal flow, simulating the turbulence driven ZFs. Generating small-scale perturbations in the vicinity of the MHD modes, without the need for high spatial accuracy, in principle can be achieved by means of Radio Frequency waves (RF). As it is already demonstrated, in fusion plasmas application of RF heating has a stabilisation impact on the core MHD modes such as Sawteeth \cite{Campbell1988,Larche2017}.       




\begin{thebibliography}{100}
\bibitem{Marshall1973} P. Marshall, N. Rosenbluth, and R. Y. Dagazian, {\it Phys. of Fluids} {\bf 16}, 1894-1902 (1973). 
\bibitem{VonGoeler1974} S. Von Goeler, W. Stodiek, and N. Sauthoff, {\it Phys. Rev. Lett.} {\bf 33(20)}, 1201-1203 (1974). 
\bibitem{Ksdomtsev1975} B. B. Kadomtsev, `` Disruptive instability in Tokamaks," {\it Fiz. Plazmy} {\bf 1}, 710-715 (1975). B. B. Kadomtsev, [{\it Sov. J. Plasma Phys.} {\bf 1}, 389-391 (1975)].
\bibitem{Thyagaraja1991a} A. Thyagaraja and F. A. Haas, {\it Phys. of Plasmas} {\bf 3 (3)}, 580 (1991).
\bibitem{Porcelli1996} F. Porcelli, D. Boucher, and M. N. Rosenbluth, {\it Plasma Phys. Controlled Fusion} {\bf 38}, 2163-2186 (1996). 
\bibitem{Monticello1986} D. A. Monticello, W. Park, R. Izzo, and K. McGuire, {\it Comput. Phys. Commun.} {\bf 43(1)}, 57-67 (1986). 
\bibitem{Mendonca2018} J. Mendonca, D. Chandra, A. Sen, and A. Thyagaraja, {\it Phys. of Plasmas} {\bf 25}, 022504 (2018).
\bibitem{Kai-Qi2018} Kai-Qi Cao and Xian-Qu Wang, {\it Matter and Radiation at Extremes} {\bf 3}, 243 (2018).
\bibitem{Wilson1996} H. Wilson, J. Connor, R. Hastie, and C. Hegna, {\it Phys. of Plasmas} {\bf 3}, 248 (1996).
\bibitem{Haye2002} R. L. Haye, S. Gunter, D. Humphreys, J. Lohr, T. Luce, M. Maraschek, C. Petty, R. Prater, J. Scoville, and E. Strait, {\it Phys. of Plasmas} {\bf 9}, 2051 (2002).
\bibitem{Smolyakov1993} A. I. Smolyakov, {\it Plasma Phys. and Controlled Fusion} {\bf 35(6)}, 657 (1993).
\bibitem{Hegna1998} C. Hegna, {\it Phys. of Plasmas} {\bf 5}, 1767 (1998).
\bibitem{ZChang1995} Z. Chang, J. Callen, E. Fredrickson, R. Budny, C. Hegna, K. Mcguire, and M. Zarnstorff, {\it Phys. Rev. Lett.} {\bf 74}, 4663 (1995).
\bibitem{Carrera1986} R. Carrera, R. Hazeltime, and M. Kotschenreuther, {\it Phys. Fluids} {\bf 29}, 899 (1986).
\bibitem{McDevitt2006} C. J. McDevitt and P. H. Diamond, {\it Phys. of Plasmas} {\bf 13}, 032302 (2006). 
\bibitem{Chandra2015} D. Chandra, A. Thyagaraja, A. Sen, C. Ham, T. Hender, R. Hastie, J. Connor, P. Kaw, and J. Mendonca, {\it Nucl. Fusion} {\bf 55(5)}, 053016 (2015).
\bibitem{Larche2017} E. Lerche, M. Lennholm, I.S. Carvalho, P. Dumortier, F. Durodie, D. Van Eester, J. Graves, P. Jacquet, A. Murari and JET Contributors, {\it Nucl. Fusion} {\bf 57}, 036027 (2017).
\bibitem{Campbell1988} D. J. Campbell, D. F. H. Start, J. A. Wesson, C. V. Bartlett, V. P. Bhatnagar, M. Bures, J. G. Cordey, G. A. Cottrell, P. A. Dupperex, A. W. Edwards, C. D. Challis, C. Gormezano, C. W. Gowers, R. S. Granetz, H. Hamnen, T. Hellsten, J. Jacquinot, E. Lazzaro, P. J. Lomas, N. Lopes Cardozo, P. Mantica, J. A. Snipes, D. Stork, P. E. Stott, P. R. Thomas, E. Thompson, K. Thomsen, and G. Tonetti, {\it Phys. Rev. Lett.} {\bf 60}, 2148 {1988}
\bibitem{McClements1996} K. G. McClements, R. O. Dendy, R. J. Hastie, and T. J. Martin, {\it Phys. of Plasmas} {\bf 3}, 2994 (1996). 
\bibitem{Yang2012} Xi. Yang, Sh. Wang, and W.Yang, {\it Phys. of Plasmas} {\bf 19}, 072503 (2012).
\bibitem{Mao2001} Jianshan Mao, P.E. Phillips, K.W. Gentle, Junyu Zhao, Xianmei Zhang, Jiarong Luo, Jiangang Li, Shouyin Zhang, Xiang Gao, Yadong Li, Yinxian Jie, Zhenwei Wu, Guangli Kuang, Liqun Hu, Shengxia Liu, Xiaodong Zhang, Ning Qiu, Xiaoning Liu, Yi Bao, Yuhong Xu, Kun Yang, Guangxin Wang, Weiwei Ye, L. Chen, Yaojiang Shi, Mei Song, P.J. Qin, Xuemao Gu, Ningzhuo Cui, Hengyu Fan, Y.F. Chen, Chengyi Xia, Huailin Ruan, Xingde Tong, Jikang Xie, {\it Nucl. Fusion}, {\bf 41}, 11, 1645 (2001).
\bibitem{DeVries1997} P. C. De Vries, A. J. H. Donne, S. H. HeijnenH, C. A. J. Hugenholtz, A. Kramer-Flecken, F. C. Schuller, G. Waidmann, {\it Nucl. Fusion} {\bf 37}, 1641 (1997).
\bibitem{DeVries2009} P. C. de Vries, M. F. Johnson, I. Segui and JET EFDA Contributors, {\it Nucl. Fusion} {\bf 49(5)}, 055011 (2009).
\bibitem{DeVries2014} P. C. de Vries, M. Baruzzo, G. M. D. Hogeweij, S. Jachmich, E. Joffrin, P. J. Lomas, G. F. Matthews, A. Murari, I. Nunes, T. P\"{u}tterich, C. Reux, J. Vega, and JET-EFDA Contributors, {\it Phys. of Plasmas} {\bf 21}, 056101 (2014).
\bibitem{Lehnen2013} M. Lehnen, G. Arnoux, S. Brezinsek, J. Flanagan, S.N. Gerasimov, N. Hartmann, T.C. Hender, A. Huber, S. Jachmich, V. Kiptily, {\it Nucl. Fusion} {\bf 53 (9)}, 093007 (2013).
\bibitem{Ishizawa2019} A. Ishizawa, Y. Kishimoto and Y. Nakamura, {\it Plasma Phys. Control. Fusion} {\bf 61}, 054006 (2019).
\bibitem{Ishizawa2015} A. Ishizawa, S. Maeyama, T-H. Watanabe, H. Sugama and N. Nakajima, {\it J. Plasma Phys.} {\bf 81}, 435810203 (2015).
\bibitem{Coppi1977} B. Coppi and C. Spight, {\it Phys. Rev. Lett.} {\bf 41}, 551 (1977).
\bibitem{Kadomtsev1971} B. B. Kadomtsev and O.P. Pogutse, {\it Nucl. Fusion} {\bf 11(1)}, 67 (1971).
\bibitem{Dorland2000} W. Dorland, F. Jenko, M. Kotschenreuther, and B. N. Rogers, {\it Phys. Rev. Lett.} {\bf 85}, 5579 (2000).
\bibitem{Pueschel2008} M. J. Pueschel, M. Kammerer and F. Jenko, {\it Phys. of Plasmas} {\bf 15}, 102310 (2008).
\bibitem{Doerk2011} H. Doerk, F. Jenko, M. J. Pueschel and D. R. Hatch, {\it Phys. Rev. Lett.} {\bf 106}, 155003 (2011).
\bibitem{Guttenfolder2012} W. Guttenfelder, J. Candy, S. M. Kaye, W. M. Nevins, E. Wang, J. Zhang, R. E. Bell, N. A. Crocker, G. W. Hammett, B. P. LeBlanc, D. R. Mikkelsen, Y. Ren, and H. Yuh, {\it Phys. of Plasmas} {\bf 19}, 022506 (2012).
\bibitem{moradi2013} S. Moradi, I. Pusztai, W. Guttenfelder, T. F\"{u}l\"{o}p and A. Moll\'{e}n, {\it Nucl. Fusion} {\bf 53 (6)}, 063025 (2013).
\bibitem{Li2012} J. Li and Y. Kishimoto, { \it Phys. of Plasmas} {\bf 19}, 030705 (2012). 
\bibitem{Ishizawa2013} A. Ishizawa and F.L. Waelbroeck, {\it Phys. of Plasmas} {\bf 20}, 122301 (2013). 
\bibitem{Sun2018} P. J. Sun, Y. D. Li, Y. Ren, X. D. Zhang, G. J. Wu, L. Q. Xu, R. Chen, Q. Li, H. L. Zhao, J. Z. Zhang, T. H. Shi, Y. M. Wang, B. Lyu , L. Q. Hu, J. Li and The EAST Team, {\it Nucl. Fusion} {\bf 58}, 016003 (2018). 
\bibitem{Li2009} J. Li, Y. Kishimoto, Y. Kouduki, Z. X. Wang and M. Janvier, {\it Nucl. Fusion} {\bf 49}, 095007 (2009).
\bibitem{poli2009} E. Poli, A. Bottino and A. G. Peeters, {\it Nucl. Fusion} {\bf 49}, 075010 (2009).
\bibitem{harriri2015} F. Hariri, P. Hill, M. Ottaviani and Y. Sarazin, {\it Plasma Phys. Control. Fusion} {\bf 57}, 054001 (2015).
\bibitem{hornsby2015} W. A. Hornsby, P. Migliano, R. Buchholz, D. Zarzoso, F. J. Casson, E. Poli and A. G. Peeters, {\it Plasma Phys. Control. Fusion} {\bf 57}, 054018 (2015). 
\bibitem{Bardoczi2017} L. Bardoczi, T. A. Carter, R. J. La Haye, T. L. Rhodes and G. R. McKee, {\it Phys. of Plasmas} {\bf 24}, 122503 (2017).
\bibitem{Bardoczi2016} L. Bardoczi, T. L. Rhodes, T. A. Carter, A. Banon Navarro, W. A. Peebles, F. Jenko and G. R. McKee, {\it Phys. Rev. Lett.} {\bf 116}, 215001 (2016).
\bibitem{Diamond2005} P. H. Diamond, S-I. Itoh, K. Itoh and T. S. Hahm, {\it Plasma Phys. Control. Fusion} {\bf47}, R35 (2005).
\bibitem{Biskamp1983} D. Biskamp and H. Welter, {\it Phys. Lett. A} {\bf 96}, 25 (1983).
\bibitem{Diamond1984} P. H. Diamond, R. D. Hazeltine, Z. G. An, B. A. Carreras and H. R. Hicks, {\it Phys. of Fluids} {\bf 27}, 1449 (1984).
\bibitem{Ham1995} T. S. Hahm and K. H. Burrell, {\it Phys. of Plasmas} {\bf 2}, 1648 (1995).
\bibitem{Ham2015} G. J. Choi and T. S. Hahm, {\it Nucl. Fusion} {\bf 55(9)}, 093026 (2015).
\bibitem{Thyagaraja1991} A. Thyagaraja, R. D. Hazeltine, and A. Y. Aydemir, {\it Physics of Fluids B: Plasma Physics} {\bf 4}, 2733 (1992).
\bibitem{Thyagaraja1993} A. Thyagaraja, and F. A. Haas, {\it Physics of Fluids B: Plasma Physics} {\bf 5}, 3252 (1993).
\bibitem{Thyagaraja2010} A. Thyagaraja, M. Valovi\v{c}, and P. J. Knight, {\it Phys. of Plasmas} {\bf 17}, 042507 (2010).
\end{thebibliography}
\end{document}